\documentclass[aps,prr,twocolumn,amsmath,amssymb,showpacs,superscriptaddress,notitlepage,longbibliography]{revtex4-2}
\usepackage{graphicx}
\usepackage{subfigure}
\usepackage{dcolumn}
\usepackage{bm}
\usepackage{amsfonts}
\usepackage{amssymb}
\usepackage{amsmath}
\usepackage{color}
\usepackage[colorlinks=true,breaklinks=true,linkcolor=blue,anchorcolor=red,citecolor=blue,urlcolor=blue]{hyperref}
\usepackage{longtable}
\usepackage{booktabs}
\usepackage{multirow}
\allowdisplaybreaks[3]

\def\l{\left}
\def\r{\right}

\newcommand{\s}{{\sigma}}

\begin{document}

\title{Emergent space-time supersymmetry at disordered quantum critical points}

\author{Xue-Jia Yu}
\affiliation{International Center for Quantum Materials, School of Physics, Peking University, Beijing 100871, China}
\author{Peng-Lu Zhao}
\affiliation{Shenzhen Institute for Quantum Science and Engineering and Department of Physics,
	Southern University of Science and Technology, Shenzhen}
\affiliation{Shenzhen Key Laboratory of Quantum Science and Engineering, Shenzhen 518055, China}
\author{Shao-Kai Jian}
\affiliation{Department of Physics, Brandeis University, Waltham, Massachusetts 02453, USA}
\author{Zhiming Pan}
\email{panzhiming@westlake.edu.cn}
\affiliation{Institute of Natural Sciences, Westlake Institute for Advanced Study, Hangzhou 310024, China}
\affiliation{Department of Physics, School of Science, Westlake University, Hangzhou 310024, China}
\date{\today}

\begin{abstract}
	We study the effect of disorder on the spacetime supersymmetry that is proposed to emerge at the quantum critical point of pair density wave transition in (2+1)D Dirac semimetals and (3+1)D Weyl semimetals. 
	In the (2+1)D Dirac semimetal, we consider three types of disorder, including random scalar potential, random vector potential and random mass potential, while the random mass disorder is absent in the (3+1)D Weyl semimetal. 
	Via a systematic renormalization group analysis, we find that any type of weak random disorder is irrelevant due to the couplings between the disorder potential and the Yukawa vertex.		
	The emergent supersymmetry is thus stable against weak random potentials. 
	Our work will pave the way for exploration supersymmetry in realistic condensed matter systems.
\end{abstract}

\maketitle


\section{Introduction}\label{Sec_intro}

About five decades ago, the spacetime supersymmetry (SUSY) was proposed as a possible way of solving the hierarchy problem of the standard model \cite{Weinbergbook, Gervais-Sakita1971, Wess-Zumino1974,Dimopoulos-Georgi1981} and the cosmological constant problem\cite{Cremmer1983}. 
Later, some supersymmetric theories have been studied as toy models to understand strong coupling physics rigorously \cite{SW, SEIBERG}. 
Due to these attractive features, SUSY has been studied intensively in past fifty years and there is some expectation before that SUSY may be revealed in the large hadron collider (LHC). 
Unfortunately, the recent experiments at the LHC have found no evidence of SUSY and/or its spontaneous breaking in particle physics.

Three-dimensional (3D) Weyl fermions \cite{ashvin2011, xu2011, burkov2011} in noncentrosymmetric materials \cite{weng2015, hasan2015, xu2015, lv2015, yang2015} provide an opportunity to test and investigate important concepts developed in the context of high-energy physics in realistic condensed matter systems.
It has been suggested that SUSY can emerge in the low-energy limit of a number of non-supersymmetric models \cite{CURCI, GOH, SCOTT,juven2021,juven2021prb,Alex2021,zixiang2018sa,zixiang2017prb,zixiang2017prl,shaokai2017prl,Yu2019prb}. 
In particular, SUSY is proposed to emerge at quantum critical points (QCPs) in Bose-Fermi lattice models \cite{lee2007, yu2010}, in the (2+1)D surface states of topological insulators 
\cite{kane2010, qi2011, grover2014,ponte2014, zerf2016}, as well as at multicritical points in some low-dimensional systems \cite{friedan1984, foda1988, huijse2015}.
Moreover, an interesting recent suggestion \cite{jian2015} is that SUSY can be realized at certain pair-density-wave (PDW) superconducting quantum critical points of ideal Weyl semimetals \cite{ruan2016a, ruan2016b}(WSMs).

The realization of SUSY at QCPs relies crucially on the fact that the infrared fixed point is stable against small perturbations. 
In particular, for the emergent SUSY to be realized, it must be robust when the fermions are subject to small perturbations from quenched disorder and other dissipation effects. 
Here, we are particularly interested in the impact of quenched disorder on the emergent SUSY, because disorder unavoidably exists in all realistic materials. 
It is well known that disorder plays an essential role in condensed matter systems~\cite{Lee85, Altshuler,Fradkin1986, Belitz94, Abrahams01, Altland02, Sarma11, Kotov12,yerzhakov2018prb}
and may lead to a plenty of prominent phenomena, such as Anderson localization and metal-insulator transition. 
In graphene-like Dirac semimetals (DSMs), depending on the specific type, disorder can either enhance or reduce the effective Coulomb interaction strength~\cite{Stauber2005PRB, Herbut08, Vafek08, Goswami2011PRL, WangLiu14,Roy2014PRB, ZhaoPRB2016}, which in turn drastically modifies the phase diagram obtained in the clean limit \cite{Stauber2005PRB, Herbut08, Vafek08, Goswami2011PRL, WangLiu14, Roy2014PRB, ZhaoPRB2016}. 
Moreover, disorder may have a significant impact on the low-temperature properties of various Dirac or Weyl semimetals, such as the conductivity of graphene \cite{Ostrovsky06, Herbut08, Foster08,yerzhakov2020npb}, the optical conductivity of WSMs \cite{Roy2016SCP}, and the low-energy spectral, thermodynamic, and transport behaviors of $d$-wave cuprate superconductors \cite{Kim97, Altland02, Lee06,JWang13, Wang2011PRB}. 
Disorder also plays a vital role in quantum Hall systems \cite{Furneaux1995, YePRL98, YePRB99, Ludwig1994} and topological insulators \cite{kane2010, qi2011}.

\begin{figure}[t]
	\centering
	\includegraphics[width=0.9\linewidth]{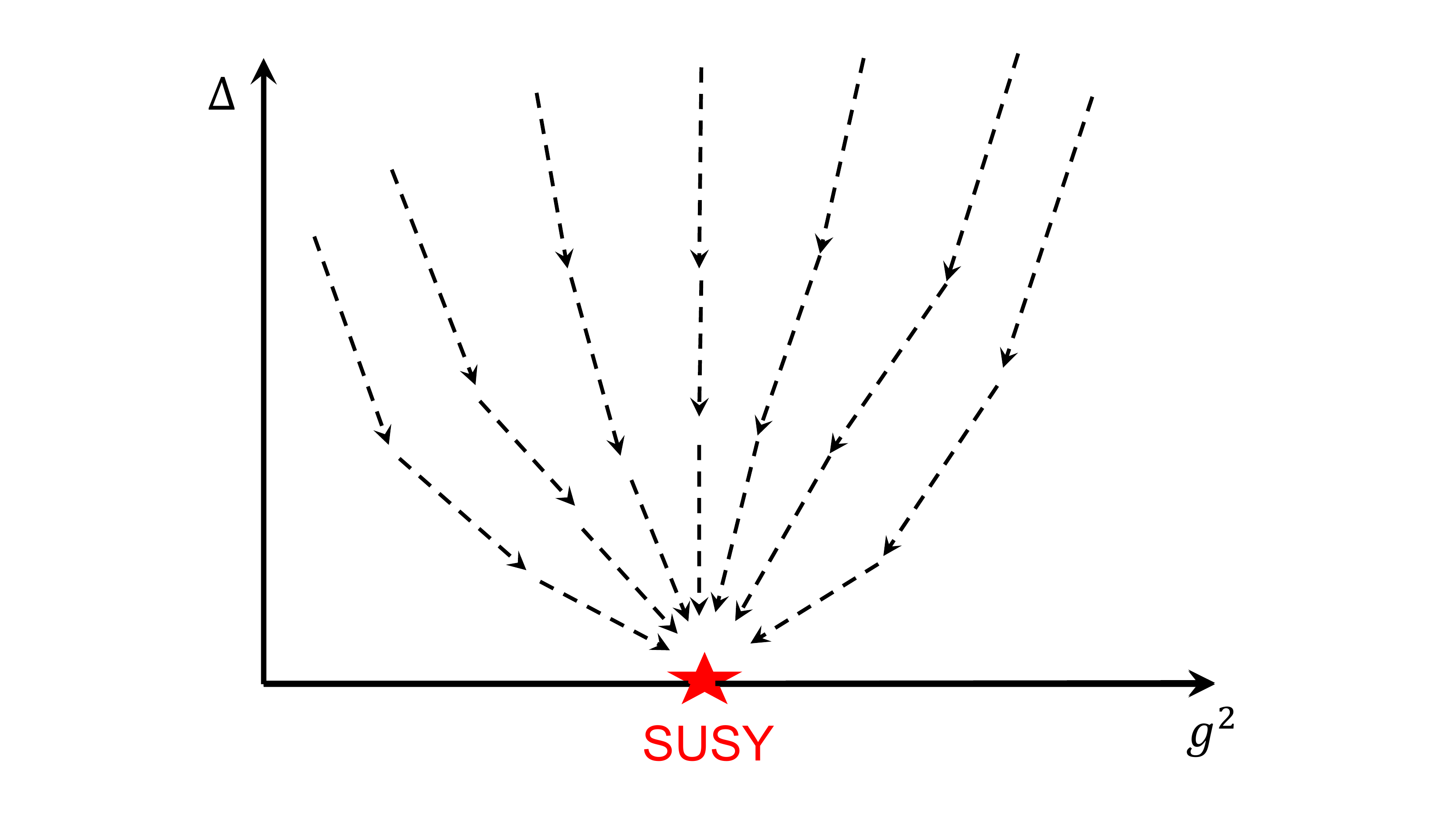}
	\caption{Schematic RG flow near SUSY fixed point as a function the Yukawa coupling $g^{2}$ and quench disorder strength $\Delta$. The emergent SUSY fixed point is stable for weak random quench disorders from our one-loop level RG calculations.}
	\label{phase_diagram}
\end{figure}

In this paper, we investigate the stability of emergent SUSY against the disorder scattering. 
We focus on the disorder-induced unusual renormalization of the fermion velocity \cite{WangLiu14, JWang13, plzhao16}, and examine whether such a renormalization effect causes a substantial difference between the velocities of fermions and bosons at low energies, and ruins the emergent SUSY. 
Based on this analysis, we are able to identify the influence of non-magnetic disorder on the \textit{particular} fixed point that is argued to display an emergent SUSY at the QCP of pair density wave (PDW) transition in (3+1)D WSMs and (2+1)D DSMs \cite{jian2015}. 
In the case of (2+1)D DSMs, we consider three types of disorder, including random scalar potential (RSP), random vector potential (RVP), and random mass (RM). 
In (2+1)D, our systematic RG analysis reveals that weak RVP, RMP and RSP are irrelevant at the QCPs where fermion velocity and boson velocity flow into same value under renormalization, which certainly does not breaks the emergent SUSY.
In (3+1)D WSMs, the disorder potential becomes more irrelevant and the effective SUSY is robust against weak disorder. 
The schematic RG flow diagram for emergent SUSY is shown in Fig.\ref{phase_diagram}

This paper is structured as follows. In Sec.~\ref{2d}, we present the effective model for (2+1)D disordered DSMs and perform the RG calculations. In Sec.~\ref{3d}, the same analysis is carried out in (3+1)D WSMs. We briefly summarize the results of this work in Sec.~\ref{Sec_summary}. Further RG details for our calculations are provided in Appendix.

\section{(2+1)D Dirac semimetals}\label{2d}

As demonstrated in Ref.~\cite{jian2015}, a spactime SUSY could emerge in the low-energy region at the PDW QCP of (2+1)D DSMs only when the number of massless Dirac fermions is $N_f = 2$. 
In this case, the low-energy effective action at the PDW criticality in a clean system is given by 
\begin{eqnarray}
	S &=& S_{f}+ S_{b} + S_I,\label{ST}\\
	S_{f} &=& \int d^2x d\tau \sum_{n=\pm} \psi_n^\dag\Big[
	\partial_\tau+iv_{f} \sum_{j=1}^2 \gamma_j
	\partial_j \Big] \psi_n\label{SF},
	\\
	S_{b} &=& \int d^2x d\tau \bigg\{\sum_{n=\pm}\Big[|\partial_\tau
	\phi_n|^2 +v_{b}^2 \sum_{j=1}^2 |\partial_j \phi_n|^2
	\\\nonumber
	&&+r|\phi_n|^2 +u|\phi_n|^4\Big]+u_{+-} |\phi_1|^2 |\phi_2|^2
	\bigg\}\label{SB},
	\\
	S_I &=& \int d^2x d\tau g \sum_{n=\pm} \Big[ \phi_n \psi_n^{T} \sigma_y
	\psi_n + h.c.\Big]\label{SI},
\end{eqnarray}
where $\gamma_j=(\sigma_x,\sigma_y)$. $\sigma_{\alpha},\alpha=x,y$ is Pauli matrix with spin indices.
$S_f$ corresponds to the action for two non-interacting two-component Dirac fermions $\psi_\pm$ at two Dirac points $\bm{Q}_{\pm}$ \cite{Gonzalez93, Gonzalez94}, with quartic and higher order self-coupling terms being irrelevant \cite{lee2007} at low energies.
$S_b$ describes the quantum fluctuation and the self-coupling of the PDW order parameter $\phi_n$ near the QCP, where $\phi_\pm$ is the superconducting order with momentum $2 \bm{Q}_{\pm}$, respectively. 
Terms with higher powers of $\phi_n$ are all irrelevant, whereas $\phi_n^{\ast}\partial_{\tau}\phi_n$ is excluded by particle-hole symmetry \cite{jian2015}. 
$S_I$ represents the Yukawa coupling between Dirac fermions and bosons. 
The terms of the form $\phi_-^{\ast} \psi_+ \s_y \psi_+$ and $\phi_-^{\ast} \psi_+\s_y \psi_-$ are not allowed because they do not satisfy momentum conservation \cite{lee2007}. 
Therefore, the effective action given above is of the most general form. 
It has been shown through renormalization group analysis that an emergent spacetime SUSY occurs at the low energy limit.
A necessary condition for the emergent SUSY is that velocities of fermions and bosons flow to the same value under RG, which renders the emergent Lorentz symmetry. 
It was claimed that such an emergent Lorentz symmetry can be naturally realized in a number of correlated electron systems~\cite{lee2007, yu2010, jian2015, kane2010, qi2011, grover2014, ponte2014, zerf2016}.

\begin{figure}[htbp]
	\subfigure{\includegraphics[width=3.0in]{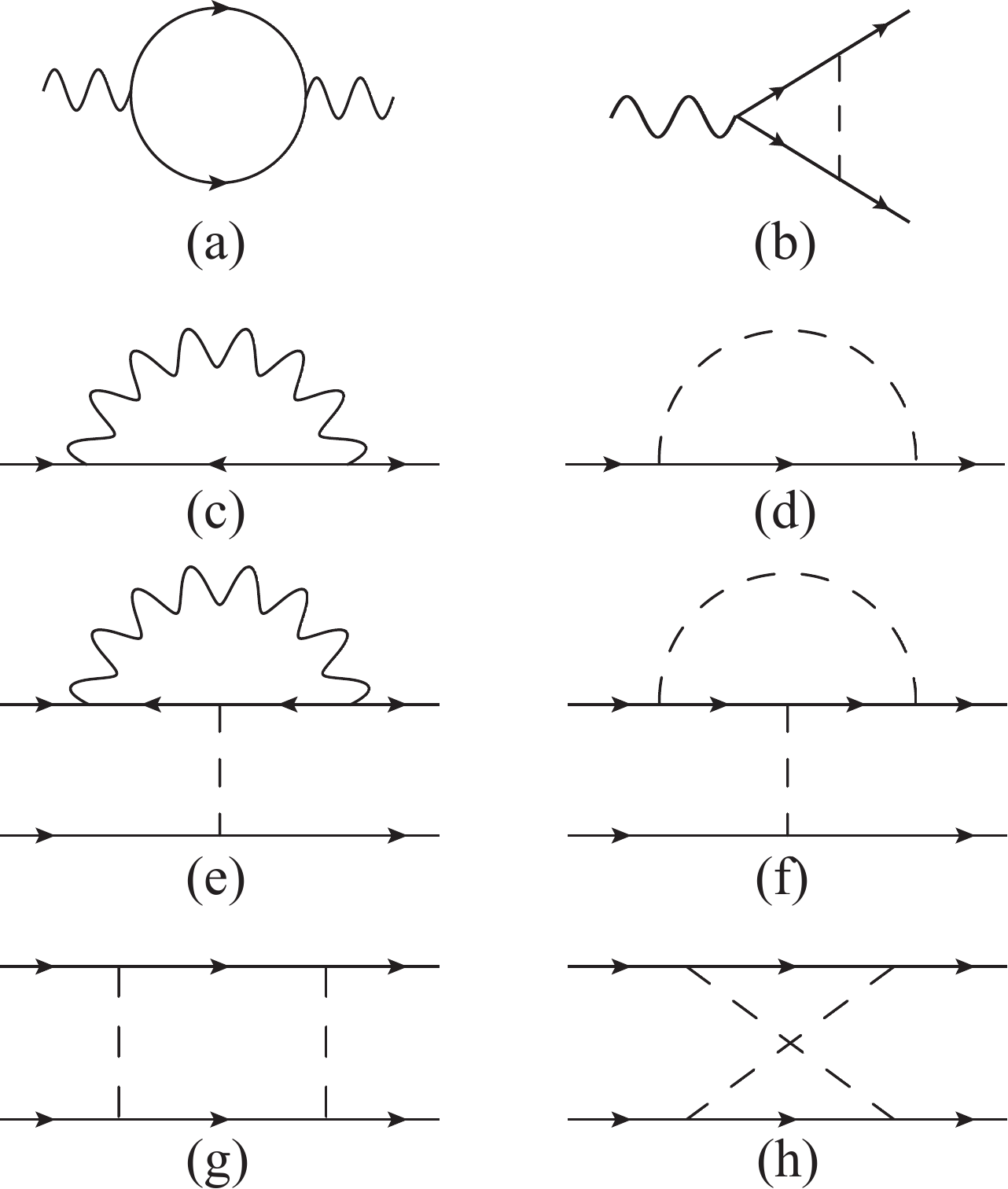}} \caption{Feynman diagrams for all
		the relevant one-loop diagrams that survive within replica limit. Here, the solid line represents the
		free fermion propagator, the wavy line the free boson propagator,
		and the dashed line the disorder.} \label{Figs}
\end{figure}

The aim of the present work is to examine whether the emergent SUSY is robust against disorder scattering. 
For this purpose, we now introduce a direct fermion-disorder coupling term to the system via
the standard form, also see Appendix A \cite{Ludwig1994, Nersesyan95, Altland02, Stauber2005PRB, Herbut08, Vafek08, Wang2011PRB, WangLiu14},
\begin{eqnarray}
	S_{\mathrm{dis}} = \int d^2x d\tau \sum_{n=\pm}\psi_n^\dag
	\left(\sum_{\Gamma} V_{\Gamma}(\mathbf{x})\Gamma\right)\psi_n,	\label{eq:Sdis}
\end{eqnarray}
where $V_{\Gamma}(\mathbf{x})$ stands for the random potential and $\Gamma$ labels the type of the disorder potential. 
We assume $V_{\Gamma}(\mathbf{x})$ to be a quenched, Gaussian white noise potential characterized by the following identities:
\begin{eqnarray}
	\langle V_{\Gamma}(\mathbf{x})\rangle = 0,\quad \langle
	V_{\Gamma}(\mathbf{x})V_{\Gamma'}(\mathbf{x}')\rangle = 
	\Delta_{\Gamma} \delta_{\Gamma \Gamma'} \delta(\mathbf{x}-\mathbf{x}'),\label{Eq_def_dis}
\end{eqnarray}
where $\langle ... \rangle$ denotes average over disorder distribution and $\Delta_{\Gamma}$ is introduced to characterize the strength of random potential. 

We consider three different types of disorder classified by the different matrices $\Gamma$. 
In particular, $\Gamma = \mathbb{I}_{2\times 2}$ for RSP, $\Gamma = \sigma_{z}$ for RM, and $\Gamma = (\sigma_{x},\sigma_{y})$ for RVP. 
These three types are most frequently studied in the literature and they can be induced by some specific mechanisms in realistic materials \cite{Nersesyan95, CastroNeto, Peres2010RMP, Mucciolo2010JPCM, Meyer2007Nature, Champel2010PRB, Kusminskiy2011PRB}. 
These three types of random potential might exist individually, or coexist in the same material. 
To be general, we assume that they coexist in the system and analyze their impact by performing RG calculations.

The random potential $V(\mathbf{x})$ can be properly averaged by employing the replica trick \cite{Lee85, Lerner, Goswami2011PRL, Roy2014PRB, Roy2016PRB, Roy2016SCP}, which leads us to an interacting effective action of short-range fermion-fermion interaction:
\begin{widetext}
	\begin{align}
		S_{\mathrm{dis}} =&-\frac{1}{2}\int d^2x d\tau
		d\tau' \Big\{ \sum_{n=\pm}\Big[
		\Delta_{S}\big(\psi_{n}^{\dagger \alpha}
		\psi_{n}^{\alpha}\big)_x \big(\psi_{n}^{\dagger \beta}
		\psi_{n}^{\beta}\big)_{x'}
		+\Delta_M\big(\psi_{n}^{\dagger \alpha}\s_z
		\psi_{n}^{\alpha}\big)_x \big(\psi_{n}^{\dagger \beta}\s_z
		\psi_{n}^{\beta}\big)_{x'}		\nonumber\\
		&+\Delta_V \sum_{j} \big(\psi_{n}^{\dagger \alpha}\s_{j}
		\psi_{n}^{\alpha}\big)_x
		\big(\psi_{n}^{\dagger \beta}\s_{j}
		\psi_{n}^{\beta}\big)_{x'}\Big]
		+2\Delta_{S}^{\prime} \big(\psi_{+}^{\dagger \alpha}
		\psi_{+}^{\alpha}\big)_x \big(\psi_{-}^{\dagger \beta}
		\psi_{-}^{\beta}\big)_{x'}		\nonumber\\
		&
		+2\Delta_M^{\prime}\big(\psi_{+}^{\dagger \alpha}\s_z
		\psi_{+}^{\alpha}\big)_x \big(\psi_{-}^{\dagger \beta}\s_z
		\psi_{-}^{\beta}\big)_{x'}
		+2\Delta_V^{\prime} \sum_{j} \big(\psi_{+}^{\dagger \alpha}\s_{j}
		\psi_{+}^{\alpha}\big)_x
		\big(\psi_{-}^{\dagger \beta}\s_{j}
		\psi_{-}^{\beta}\big)_{x'}
		\Big\} 
		\label{SD}
	\end{align}
\end{widetext}
where $j=(x,y)$, $\alpha$ and $\beta$ are the replica indices, $x \equiv(\mathbf{x},\tau)$ and $x'\equiv(\mathbf{x},\tau')$ are the space-time coordinate. 
The repeated indices $\alpha$ and $\beta$ are summed automatically.
In the replica theory, the replica limit $\sum_{\alpha}=N\rightarrow 0$ is implemented in the following RG calculation. 
Three parameters $\Delta_S$, $\Delta_M$, and $\Delta_V$ characterize the effective strength of quartic
couplings of Dirac fermions induced by averaging over RS, RM, and
RVP, respectively.
The two pieces of Dirac fermions share the same random potential.
Three cross terms characterized by $\Delta_S^{\prime}$, $\Delta_M^{\prime}$, and $\Delta_V^{\prime}$ are induced in the replica limit. 
In the RG analysis, the bare values of the parameters are the same, $\Delta_S^{\prime0}=\Delta_S^{0}$, $\Delta_M^{\prime0}=\Delta_M^{0}$, and $\Delta_V^{\prime0}=\Delta_V^{0}$, moreover, the RG equations for $\Delta_\Gamma$ and $\Delta_{\Gamma'}$ are the same (see Appendix B for details), so we focus on $\Delta_\Gamma$ in the following. 

As shown in previous calculations \cite{WangLiu14, JWang13, plzhao16}, disorder can strongly affect the RG flow of fermion velocity as the energy is lowered. 
If the disorder coupling is relevant that flow to a finite value at low energy limit, it will drive the fermion velocity to vanish at sufficiently low energies, which then spoils the Lorentz symmetry for the fermion sector, but not for the boson sector. 
As a result, the emergent Lorentz symmetry, and thus the emergent SUSY will be ruined by disorder. 
However, whether this takes place relies crucially on the scale dependence of disorder coupling parameter. 
In the case of (2+1)D DSMs, naive power-counting, according to Eqs.~(\ref{SF}) and (\ref{SD}), shows that disorder is marginal. 
A careful analysis of the marginal disorder effect is helpful to tell us whether a irrelevant and a relevant coupling need to investigate further.

To this end, we carry out a detailed RG analysis starting from the critical action with $r=0$, represented by Eq.~(\ref{ST}) along with Eq.~(\ref{SD}), by considering the leading order of the $\epsilon$-expansion, where $\epsilon = 4-D=3-d$, $D$ and $d$ are the spacetime dimension and the spatial dimension, respectively. 
The pertinent one-loop Feynman diagrams are shown in Fig.~\ref{Figs}. 
After integrating out the fast modes defined within the momentum shell $e^{-l}\Lambda < |\mathbf{p}| < \Lambda$ and then performing RG transformations \cite{Shankar1994RMP}, we obtain the following RG equations (the detailed results are presented in appendix)
\begin{widetext}
	\begin{align}
		\frac{d v_f}{dl}
		&=v_{f} \Big[ g^2 (G_1 -G_0) -2\sum_{\Gamma} \Delta_{\Gamma}  \Big],	    \label{RGvf}\\
		\frac{da}{dl}
		&=g^2 \Big( \frac{1 -a^2}{2a} 
		+ a(G_0 -G_1) \Big) +2a \sum_{\Gamma} \Delta_{\Gamma},   \label{RGa}\\
		\frac{dg^2}{dl}
		&=\epsilon g^2  -g^4 \big(1 -G_0 +3G_1\big) +\big( 4\Delta_S +2\sum_{\Gamma} \Delta_{\Gamma} \big) g^2,   \label{RGg^2}\\
		\frac{d\Delta_S}{dl} 
		&= \big( \epsilon-1 \big) \Delta_S +2\Delta_S \big( \Delta_S+ \Delta_M +2 \Delta_V \big) +4\Delta_{M} \Delta_{V} +(G_0- 3G_2 -2G_1) \Delta_S g^2,    \label{RGS}\\
		\frac{d\Delta_M}{dl} 
		&= 
		\big( \epsilon-1  \big) \Delta_M -2\Delta_{M} \big( \Delta_S+ \Delta_M -2 \Delta_V \big) +4\Delta_S \Delta_{V} -(G_0+ 3G_2+2G_1) \Delta_M g^2,    \label{RGM}\\
		\frac{d\Delta_V}{dl}
		&= \big( \epsilon-1 \big) \Delta_V+
		2\Delta_{M} \Delta_S -(G_0+2G_1) \Delta_V g^2,     \label{RGV}
	\end{align}
\end{widetext}
where $a = v_b/v_f$, $G_0=\frac{4}{a(a+1)^2}$, $G_1=\frac{4(2a+1)}{3a(a+1)^2}$, and $G_2=\frac{4(2+a)}{3a(a+1)^2}$. 
In the above calculations, we have rescaled all the couplings as follows: $g^2\Lambda^{-\epsilon}S_{D-1}/[2(2\pi)^{D-1}v_f^{D-1}] \rightarrow
g^2,$ and $\Delta_{\Gamma}\Lambda^{1-\epsilon}S_{D-1}/[(2\pi)^{D-1}v_f^2] \rightarrow \Delta_{\Gamma}$, with $S_{d}=2\pi^{d/2}/\Gamma(d/2)$ as the area of the unit sphere in $d$ dimensions. 
By setting all $\Delta_{\Gamma}=0$, Eqs.~(9)~-~(10) recover the RG equations for $v_f$, $a$, and $g^2$ previously obtained in Refs.~\cite{lee2007} and \cite{jian2015}. 
In the case of disordered Dirac fermion systems with $g=0$, our RG results for $\Delta_{\Gamma}$ are in accordance with that previously obtained in~\cite{Ostrovsky06, Foster08, FosterPRB2012}. 
The RG equations for $u$ and $u_{+-}$, which are not shown here, are exactly the same as those presented in Refs.~\cite{lee2007} and \cite{jian2015}, since there is no direct coupling between boson and fermion disorder potential. 
In the case of clean system, as demonstrated in Refs.~\cite{lee2007} and \cite{jian2015}, $a^{\ast}=1$ is the only stable infrared fixed point for $a$, which means that the bosons and fermions have the same velocity at low energies. 
Moreover, the coupling constant $g$, $u$ and $u_{+-}$ will flow to a strongly coupled fixed point that preserves SUSY. 
In the following, we first analyze the effects of single disorder, and then consider the interplay between different types of disorder.

We now consider the case in which RVP exists by itself by taking $\Delta_M = \Delta_S = 0$. 
Noting that the physical case of (2+1)D corresponds to $\epsilon \rightarrow 1$, Eq.~(\ref{RGV}) becomes 
\begin{align} 
	\frac{d \Delta_V}{dl}=& -(G_0+2G_1) \Delta_V g^2 \label{DDV}
\end{align} 
Thus the effective coupling strength for RVP, namely $\Delta_V$, is irrelevant and flows to zero. 
Without the fermion-boson coupling $g$, RVP is marginal, which originates from the existence of a time-independent gauge transformation that ensures RVP unrenormalized and is valid at any order of loop expansion \cite{Ludwig1994, Herbut08, Vafek08, FosterPRB2012,plzhao16}. 
Nevertheless, near the emergent SUSY fixed point, where the coupling constant $g$ remains finite, RVP is irrelevant, and thus the emergent SUSY is stable. 

We then assume that RM exists alone, which means $\Delta_S =
\Delta_V = 0$ in Eq.~(\ref{RGM}), and we have
\begin{eqnarray}
	\frac{d\Delta_M}{dl} 
	&=& -2\Delta_{M}^2-(G_0+ 3G_2+2G_1) \Delta_M g^2.
\end{eqnarray}
From the RG function, we see that ${\Delta}_M$ is always irrelevant.
We thus can infer that the emergent SUSY is also robust again RM. 

\begin{figure}[t]
	\centering
	\includegraphics[width=0.7\linewidth]{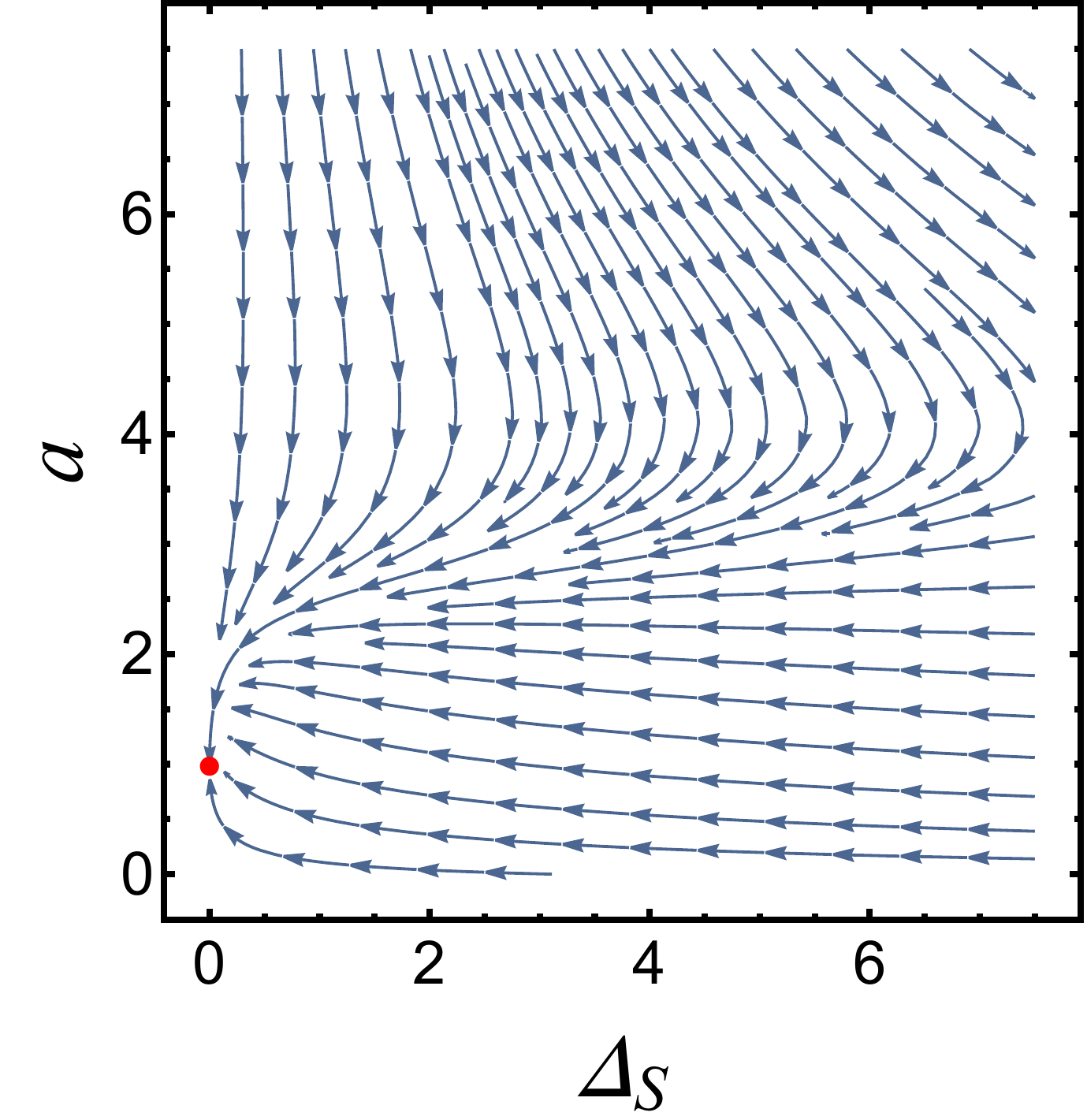}
	\caption{Flow diagram in the $\Delta_S-a$ plane with $g^2=g_c^2$ when the system contains only RSP. There is an stable fixed point(red point)	$(g^2,\Delta_S^{\ast}, a^{\ast}) = (g_c^2,0,1)$. Within $g_c^2$ plane, no other fixed point exists and any RSP $\Delta_S$ is irrelevant at one-loop level. The system always flows into weak disorder regime.}
	\label{2dRSP}
\end{figure}

The RSP can be similarly analyzed. The simplified RG equations
for RSP are
\begin{align}
	\frac{da}{dl}
	=&g^2 \Big( \frac{1 -a^2}{2a} 
	+ a(G_0 -G_1)  \Big)
	+2a \Delta_{S},  \\
	\frac{dg^2}{dl}
	=& g^2  -g^4 \big(1 -G_0 +3G_1\big) +6g^2 \Delta_S,  \\
	\frac{d\Delta_S}{dl} 
	=& 2\Delta_S^2 +(G_0- 3G_2 -2G_1) \Delta_S g^2. \label{RGSS}
\end{align}
There exist two fixed points, Gaussian fixed point ($g^2=\Delta_s=0$) and the SUSY fixed point.
The Gaussian fixed point is unstable.
With only random scalar potential ($\Delta_s\neq 0, g^2=0$), the system will flow into strong disorder regime and explicitly breaks the Lorentz symmetry.
On the contrary, $g^2$ has a finite critical point $g_c^2$,
\begin{align}
g_c^2 =\frac{1+6\Delta_S}{1 -G_0 +3G_1},
\end{align}
where the RG function of $\Delta_S$ near $\Delta_S=0$ is negative ($G_0-3G_2-2G_1<0$), and we expect small $\Delta_S$ is irrelevant.
The RG flow of the equations within the critical plane $g^2=g_c^2$ is shown in Fig.~\ref{2dRSP}.
Finite $g^2$ will always pull back the RG flow and RSP $\Delta_S$ is irrelevant eventually at one-loop level for any initial values.
We find that SUSY is stable: an arbitrarily
strong RSP flows to the weak coupling regime in the lowest energy
limit. This means that the RSP is an irrelevant perturbation, which is consistent with the one-loop RG result in \cite{Nandkishore13}. 
Previous results show that the RSP is marginally relevant for Dirac fermions, and will induce an instability of the system, leading to a diffusive motion of the Dirac fermions \cite{Fradkin1986, Altland02, Ostrovsky06, Foster08, Goswami2011PRL, FosterPRB2012, Roy2014PRB}.
However, our result shows that, the RSP is rendered irrelevant by the critical fluctuations at the PDW QCP through the finite Yukawa coupling $g^2$ between fermion and boson. 
There exists no diffusive behaviors as long as $g^2$ is finite.


When more than one type of disorder exist, from the properties of the RG equation, Eqs.~(\ref{RGM})-(\ref{DDV}) that the coexistence of any two types
disorder dynamically generate the third one. 
Thus we need to analyze the full set of RG equations given by Eqs.~(\ref{RGvf})-(\ref{RGV}). 
It is easy to see that there is a fixed point given by $(v_f, a^*,g^{2\ast}, \Delta_S^\ast, \Delta_M^\ast, \Delta_V^\ast) = (v_f, 1, \frac{\epsilon}3, 0 ,0, 0)$, where $v_f$ can take any values.
Combining with the RG equation for coupling constants $u$ and $u_{+-}$, it turns out that this is the SUSY fixed point. 
Now we examine whether this fixed point is stable by expanding the RG equation at the fixed point, and calculate eigenvalues of the stability matrix (see Appendix C).
The eigenvalues of the stability matrix are all negative except for one marginal direction at $v_f$. 
So we can conclude that the emergent SUSY is robust against weak disorders. 

\section{(3+1)D Weyl semimetals}\label{3d}

In this section, we examine the disorder effects on the emergent SUSY in (3+1)D WSMs~\cite{jian2015}.
The effective action of the disordered system in the vicinity of PDW QCP is given by
\begin{eqnarray} 
	S &=&S_{f}+ S_{b} + S_I + S_\text{dis},\label{effaction4D}
	\\
	S_{f} &=& \int d^4x \sum_{n=\pm} \Big[ \psi_n^\dag \partial_\tau
	\psi_n+\sum_{j=1}^3 i v_{fj}\psi_n^\dag \gamma_n^j \partial_j
	\psi_n\Big], \qquad
	\\
	S_{b} &=& \int d^4x \bigg\{ \sum_{n=\pm} \Big[|\partial_\tau
	\phi_n|^2 + \sum_{j=1}^3 v_{bj}^2 |\partial_j \phi_n|^2
	\\\nonumber
	&&+r|\phi_n|^2 +u |\phi_n|^4 \Big]+u_{+-} |\phi_+|^2 |\phi_-|^2
	\bigg\},
	\\
	S_I &=& \int d^4x g \sum_{n=\pm} \Big[ \phi_n \psi_n \sigma^y \psi_n
	+ h.c.\Big],
	\\
	S_\text{dis}&=& - \frac{1}{2}\int d^3x d\tau
	d\tau' \Big[\Delta_S\big(\psi^{\dagger}_{\alpha}\s_0
	\psi_{\alpha}\big)_x \big(\psi^{\dagger}_{\beta}\s_0
	\psi_{\beta}\big)_{x'}
	\\\nonumber
	&&+\sum_{i=x,y,z}\Delta_i\big(\psi^{\dagger}_{\alpha}\s_i
	\psi_{\alpha}\big)_x \big(\psi^{\dagger}_{\beta}\s_i
	\psi_{\beta}\big)_{x'}\Big], \end{eqnarray} 
where $\gamma_{\pm}^j=(\s^x,\s^y, \pm \s^z)$. 
Now, $\psi_\pm$ denotes the two-component Weyl fermions at two Weyl points $\bm{Q}_{\pm}$, and $\phi_\pm$ is the superconducting order with momentum $2 \bm{Q}_{\pm}$, respectively. 
In (3+1)D Weyl semimetal, RVP has three components, i.e., RM becomes the third component~\cite{Sbierski16, Syzranov}. 
In general, the fermion velocity is anisotropic, with unequal values along different directions. 
As shown in Ref.~\cite{jian2015}, even in the extremely anisotropic case, an emergent Lorentz symmetry can be established in the lowest energy limit. 
Our current concern is whether this emergent Lorentz symmetry can be broken by quenched disorder, thus it suffices to consider the isotopic case. 
We now can assume that $v_{fx} = v_{fy} = v_{fz} = v_{f}$, and also make the same assumption for the bosonic field. We employ the same symbol $a$ to identify the ratio between boson and fermion, and the same definition of $G_0$, $G_1$ and the rescaled couplings in Sec.~\ref{2d}. 
Calculating the same diagrams in Fig.~\ref{Figs}, we obtain the following RG equations:
\begin{widetext}
	\begin{eqnarray}
		\frac{d\ln v_{f}}{dl}&=&g^2(G_1-G_0)-2\l(\Delta_{S}+\sum_{i}\Delta_i\r)\label{RGvf3d},
		\\
		\frac{d\ln a^2}{dl}&=&\frac{g^2(1-a^2)}{a^2}
		-2g^2(G_1-G_0)+4\l(\Delta_{S}+\sum_{i}\Delta_i\r)\label{RGa3d},
		\\
		\frac{d\ln g^2}{dl}&=&\epsilon -g^2(3G_1-G_0+1)
		+\l(6\Delta_{S} +2\sum_{i}\Delta_i\r)\label{RGg^23d},
		\\
		\frac{d\Delta_S}{dl}&=&(\epsilon-1)\Delta_S+
		2\Delta_S\l(\Delta_{S}+\sum_{i}\Delta_i\r)
		+\frac{2}{3}\sum_{i}\sum_{j\neq i}\Delta_i\Delta_{j} +g^2(G_0-3G_2-2G_1)\Delta_S
		\label{RGS0},
		\\
		\frac{d\Delta_i}{dl}&=&(\epsilon-1)\Delta_i+\frac{4}{3}\sum_{j\neq
			i} \Delta_S\Delta_{j}-\frac{2}{3}\Delta_i\l(\Delta_{S} +
		2\Delta_i-\sum_{j}\Delta_{j}\r)
		-\Delta_{i} g^{2}(G_{0}+ G_{2}+2G_1)\label{RGVi},
	\end{eqnarray}
\end{widetext}
in these equations, the index $i$ is summed over $x,y,z$. The
analysis of these RG equations follows similarly as done in
Sec.~\ref{2d}. Firstly, we consider only a single vector component
disorder exists, which means 
\begin{eqnarray} 
	\Delta_i\neq 0 , \quad
	\Delta_{j\neq i}=0, \quad \Delta_{S}=0 \label{condi}, 
\end{eqnarray} 
by substituting this conditions to Eqs.~(\ref{RGvf3d}-\ref{RGVi}), three simplified RG equations for $a$, $g^2$ and
$\Delta_i$ are obtained, we just exhibit the result for disorder
coupling as 
\begin{align} 
\frac{d\Delta_i}{dl} &=
(\epsilon-1)\Delta_i-\frac{2}{3}\Delta_i^2-\Delta_{i} g^{2}(G_{0}+ G_{2}+2G_1)\label{RGsimple}.
\end{align}
Therefore, for an exact (3+1)D system corresponding to $\epsilon =
0$, any component of RVP is irrelevant. 
As a result, in (3+1)D WSMs, the emergent SUSY is robust against any
single component of RVP.

For there is only RSP in the system, we have $\Delta_i = 0$. Now
Eqs.~(\ref{RGvf3d})~-~(\ref{RGVi}) are simplified to 
\begin{align}
	\frac{d\ln a^2}{dl}=&\frac{g^2(1-a^2)}{a^2}
	-2g^2(G_1-G_0)+4\Delta_{S}\label{RGa3dS},
	\\
	\frac{d\ln g^2}{dl}=&\epsilon -g^2(3G_1-G_0+1)
	+6\Delta_{S}\label{RGg^23dS},
	\\
	\frac{d\Delta_S}{dl}=&(\epsilon-1)\Delta_S+
	2\Delta_S^2     \notag \\
	&\qquad+g^2(G_0-3G_2-2G_1)\Delta_S\label{RGS3dS}, 
\end{align}
according to Eq.~(\ref{RGg^23dS}), and noting the fact $3G_1 - G_0 + 1 > 0$, the stable fixed point for $g^2$ is located at $g_c^2=6\Delta_{S}/(3G_1 - G_0 +1)$. 
Weak RSP itself is irrelevant as indicated in \label{RGS3dS} because $G_0-3G_2-2G_1<0$. 
Whereas for strong RSP, the scenario is similar to the case of (2+1) DSMs, namely, the RSP becomes irrelevant due to the interplay between the Weyl fermion and the PDW order parameter. 
We thus conclude that the the emergent SUSY is stable against RSP.

Next, we consider the coexisting case. According to Eq.~(\ref{RGS0}), the coexistence of two components of RVP can dynamically generate RSP even when RSP does not exist at the beginning. 
We also learn from Eq.~(\ref{RGVi}) that the coexistence of RSP and any component of RVP produce the other two components.
Therefore, we need to consider the generic case in which all three components of RVP coexist with RSP. 
Now the disorder effects should be analyzed by solving the complete set of equations given by Eq.~(\ref{RGvf3d})~-~(\ref{RGVi}). 
It is hard to solve these coupled equations, but fortunately, it is simple to show that in the weak disorder regime the SUSY fixed point is stable against all random potentials, similar to the case of (2+1) DSMs.

\section{Summary and discussion}\label{Sec_summary}

In summary, we perform a standard perturbative RG to study the effect of disorder on the emergent SUSY in (2+1)D DSMs \cite{jian2015,lee2007} and (3+1)D WSMs
\cite{jian2015}. 
According to our RG results, the effective SUSY fixed point is robust against any weak disorder irrespective of the type of disorder potentials. 
Our RG analysis of the disorder effects on the emergent SUSY appearing in (2+1)D DSMs \cite{jian2015, lee2007} and (3+1)D WSMs \cite{jian2015} can be
directly extended to other analogous models, which may be more realistic to detect emergent SUSY in quantum materials.
In our one-loop RG analysis, we have omitted new vertex that could be generated from the disorder potential and the Yukawa coupling.
It will be interesting to include their effects although at tree-level they are irrelevant. 
Our work can shed new light on the understanding and exploring the emergent SUSY in realistic condensed matter systems.

\noindent
\begin{acknowledgments}
	This work was supported by National Natural Science Foundation of China (No. 12147104). The work of SKJ is also supported in part by the Simons Foundation via the It From Qubit Collaboration.
\end{acknowledgments}

\section*{Appendix A: Disorder potentials}
In this section, we briefly consider the possible disorder potential in the two-component 
Dirac (Weyl) system. The original spinful fermion annihilation operator $\psi$ can be expand around
two Dirac (Weyl) point $\bm{Q}_{\pm}$,
\begin{align}
	\psi(\bm{x}) =e^{-i\bm{Q}_+\cdot\bm{x}} \psi_{+}(\bm{x})
	+e^{-i\bm{Q}_-\cdot\bm{x}} \psi_{-}(\bm{x}),	\label{eq:psiexpanded}
\end{align}
where $\psi_{\pm}$ corresponds to the two low-energy Dirac (Weyl) fermion.
For the fermion system $\psi$, a general disorder potential takes the form,
\begin{align}
	H_{\mathrm{dis}} = \int d^d\bm{x} \psi^\dag(\bm{x})
	\left(\sum_{\Gamma} V_{\Gamma}(\mathbf{x})\Gamma\right)\psi(\bm{x}),
\end{align}
where the $V_{\Gamma}(\mathbf{x})$ stands for the randomly
distributed potential and $\Gamma=I,\sigma_{\mu}$ ($\mu=x,y,z$) labels type of the disorder potential. 
We focus on the quenched disorder potential $V_{\Gamma}(\mathbf{x})$, with Gaussian white noise potential characterized by the following identities:
\begin{eqnarray}
	\langle V_{\Gamma}(\mathbf{x})\rangle = 0,\, \langle
	V_{\Gamma}(\mathbf{x})V_{\Gamma^{'}}(\mathbf{x}')\rangle =
	\Delta_{\Gamma}\delta_{\Gamma\Gamma^{'}} \delta^d(\mathbf{x}-\mathbf{x}').
\end{eqnarray}
Substituting the fermion field Eq.~(\ref{eq:psiexpanded}) into the disordered Hamiltonian $H_{\text{dis}}$,
we can arrive at the disorder Hamiltonian for the two low-energy Dirac (Weyl) fermion $\psi_{\pm}$.
Notice that for the coupling between the two Dirac field, e.g. $\psi_+^{\dagger}(\bm{x}) \Gamma \psi_{-}(\bm{x})$, there is an overall oscillating factor $e^{\pm i(\bm{Q}_++\bm{Q}_-)\cdot \bm{x}}$.
Such disorder potential coupled the two Dirac (Weyl) fermion $\psi_{\pm}$ will be small in general.
We only need to consider the disorder Hamiltonian Eq.~(\ref{eq:Sdis}) with respect to a piece of Dirac fermion.
It should be emphasized that the two pieces of Dirac (Weyl) fermion share the same statistical distribution of random potential.

\section*{Appendix B: RG details}
We present here the detailed calculation of
Fig.~\ref{Figs}(a)-Fig.~\ref{Figs}(h) as well as the RG equations in
$(2+1)$D, the calculation for $(3+1)$D is directly followed, which
is not detailed shown here.

From the free action of fermions and bosons,
\begin{align}
	S_{f0} &= \int d^2x d\tau \sum_{n=\pm} \psi_n^\dag\Big[
	\partial_\tau+iv_{f} \sum_{j=1}^2 \gamma_j
	\partial_j \Big] \psi_n\label{SF0},
	\\
	S_{b0} &= \int d^2x d\tau \bigg\{\sum_{n=\pm}\Big[|\partial_\tau
	\phi_n|^2 +v_{b}^2 \sum_{j=1}^2 |\partial_j \phi_n|^2
	\bigg\}\label{SB0},
\end{align}
the free propagator for fermions
and bosons are 
\begin{align*}
	G_0(k) &= \frac{1}{ik_{\tau}- v_{f} \bm{\gamma}\cdot\mathbf{k}};\\
	D_0(k) &= \frac{1}{k_{\tau}^2+v_{b}^2\mathbf{k}^2}.
\end{align*}
with $k=(k_{\tau},\bm{k})$ in the momentum space through replacement $(\partial_{\tau},\partial_i) \rightarrow (ik_{\tau},ik_i)$.
The free propagators for the two pieces of Dirac fermion take the same form.
For Fig.~\ref{Figs}(a), it corresponds
\begin{align}
	\Pi(k)&= -2g^2
	\int_p \text{Tr}\left[\sigma^y G_0^T(p) \sigma^y G_0(-p-k)\right] \notag\\
	&=\frac{g^2S_d\Lambda^{-\epsilon}}{2(2\pi)^d v_f^3} l 
	\Big(k_{\tau}^2+ \Big(2 -\frac{3}{d} \Big) v_f^2\bm{k}^2\Big)
\end{align}
where $\int_{p}\equiv \int dp_{\tau}d^{d}\bm{p}/(2\pi)^D$ is the $D=d+1$ dimensional momentum integral and
$S_{d}=2\pi^{d/2}/\Gamma(d/2)$ is the area of the unit sphere in $d$
dimensions.
For Fig.~\ref{Figs}(c), it gives 
\begin{align}
	&\Sigma_{f}^{(b)}(k) =-4g^2 (-)
	\int_p \sigma^y G_0^T(p)
	\sigma^y D_0(-k-p)\nonumber\\
	&=\frac{g^2S_{d}\Lambda^{-\epsilon}}{2(2\pi)^{d}v_f^{3}} l
	\Big( \frac{4(ik_{\tau})}{a(a+1)^2}
	-\frac{4(2a+1)}{ad(a+1)^2}(v_f\bm{\gamma}\cdot\mathbf{k}) \Big)\nonumber\\
	&=\frac{g^2S_{d}\Lambda^{-\epsilon}}{2(2\pi)^{d}v_f^{3}} l
	\Big( G_0 (ik_{\tau})
	-G_1(v_f\bm{\gamma}\cdot\mathbf{k}) \Big)
\end{align}
The diagram of Fig.~\ref{Figs}(d) is ,
\begin{align}
	\Sigma_{f}^{(c)}(k)&=
	-\sum_{\Gamma}(\Delta_{\Gamma}+\Delta_{\Gamma}^{\prime})\int\frac{d^{d}\mathbf{k}}{(2\pi)^{d}}\Gamma
	G_{0}(k)\Gamma\nonumber\\
	&= \sum_{\Gamma} (\Delta_{\Gamma}+\Delta_{\Gamma}^{\prime}) \frac{\Lambda^{1-\epsilon}
		S_{d}}{(2\pi)^{d}v_f^2}(ik_{\tau})\ell,
\end{align}
which only contribute to the velocity renormalization at one-loop.
The diagram of Fig.~\ref{Figs}(e) is given by
\begin{align}
	\delta\Delta_{\Gamma}^{(d)}
	= &-8\Delta_{\Gamma} {g}^2 \int_k \sigma^y D(-k) G^T(k) \Gamma^T G^T(k) \sigma^y \\
	\delta\Delta_{\Gamma}^{\prime(d)}
	= &-8\Delta_{\Gamma}^{\prime} {g}^2 \int_k \sigma^y D(-k) G^T(k) \Gamma^T G^T(k) \sigma^y 
\end{align}
Calculating out these integrals for different $\Gamma_a$ one by one, we have
\begin{align}
	\delta\Delta_{S}^{(d)}
	=&\Delta_{S} \frac{g^2S_d\Lambda^{-\epsilon}}{2(2\pi)^d v_f^3 }
	( G_0 -dG_2 ),	\\
	\delta\Delta_{M}^{(d)}
	=&\Delta_{M} \frac{g^2S_d\Lambda^{-\epsilon}}{2(2\pi)^d v_f^3 }  ( -G_0 -dG_2 ) ,	\\
	\delta\Delta_{V}^{(d)}
	=&\Delta_{V} \frac{g^2S_d\Lambda^{-\epsilon}}{2(2\pi)^d v_f^3 } ( -G_0 ), 
\end{align}
and same for $\Delta\rightarrow \Delta^{\prime}$.
The diagram of Fig.~\ref{Figs}(b) involves $\Delta_{\gamma}$-vertex and is given by
\begin{align}
	&\delta g^{(h)}
	=g\sum_{\Gamma} \Delta_{\Gamma} \int_{\bm{k}} 
	\Gamma^T G_+^T(k) \sigma^y G_+(-k)  \Gamma	\notag\\
	=&gl\sum_{\Gamma} \frac{\Delta_{\Gamma}\Lambda^{1-\epsilon}S_{d}}{(2\pi)^{d}v_f^2} \Gamma^T \sigma^y \Gamma
\end{align} 
The remaining diagrams Fig.~\ref{Figs}(f,g,h) correspond to pure coupling between disorder potentials \cite{Ludwig1994}, which can be obtained as follows,
\begin{widetext}
	\begin{align}
		\delta\Delta_S^{(e,f,g)} =& \Big( \frac{S_d}{(2\pi)^d v_f^2 \Lambda^{2-d}} \Big) l
		\Big[ +2\Delta_{S} \big( \Delta_S+ \Delta_M +2 \Delta_V \big) +4\Delta_{M} \Delta_{V} \Big]    \\
		\delta\Delta_M^{(e,f,g)} =& \Big( \frac{S_d}{(2\pi)^d v_f^2 \Lambda^{2-d}} \Big) l
		\Big[ -2\Delta_{M} \big( \Delta_S+ \Delta_M -2 \Delta_V \big) +4\Delta_{S} \Delta_{V} \Big]    \\
		\delta\Delta_V^{(e,f,g)}=& \Big( \frac{S_d}{(2\pi)^d v_f^2 \Lambda^{2-d}} \Big) l
		\Big[ 2\Delta_{M} \Delta_{S} \Big]
	\end{align}
	and similarly for the other three terms,
	\begin{align}
		\delta\Delta_S^{\prime(e,f,g)} =& \Big( \frac{S_d}{(2\pi)^d v_f^2 \Lambda^{2-d}} \Big) l
		\Big[ +2\Delta_{S}^{\prime} \big( \Delta_S+ \Delta_M +2 \Delta_V \big) +4\Delta_{M}^{\prime} \Delta_{V}^{\prime} \Big]    \\
		\delta\Delta_M^{\prime(e,f,g)} =& \Big( \frac{S_d}{(2\pi)^d v_f^2 \Lambda^{2-d}} \Big) l
		\Big[ -2\Delta_{M}^{\prime} \big( \Delta_S+ \Delta_M -2 \Delta_V \big) +4\Delta_{S}^{\prime} \Delta_{V}^{\prime} \Big]    \\
		\delta\Delta_V^{\prime(e,f,g)}=& \Big( \frac{S_d}{(2\pi)^d v_f^2 \Lambda^{2-d}} \Big) l
		\Big[ 2\Delta_{M}^{\prime} \Delta_{S}^{\prime} \Big]
	\end{align}
	According to above results, label and rescale the couplings as follows:
	\begin{align}
		G_0=\frac{4}{a (1 +a)^2},\quad
		G_1=\frac{4(1+2a)}{d a (1 +a)^2},\quad
		G_2=\frac{4(2+a)}{da (1 +a)^2},\quad
		\frac{g^2\Lambda^{-\epsilon}S_{d}}{2(2\pi)^{d}v_f^{3}} \rightarrow
		g^2, \quad
		\frac{\Delta_{\Gamma}\Lambda^{1-\epsilon}S_{d}}{(2\pi)^{d}v_f^2}
		\rightarrow \Delta_{\Gamma}.
	\end{align}
	Then, the results obtained for Fig.~\ref{Figs}(a)-Fig.~\ref{Figs}(h) can be simplified, leading to the one-loop quantum corrections of the action.
	These one-loop results produce the RG equations 
	\begin{widetext}
		\begin{align}
			\frac{d v_f}{dl}
			&=v_{f} \Big[ g^2 (G_1 -G_0) -\sum_{\Gamma} \Delta_{\Gamma} -\sum_{\Gamma} \Delta_{\Gamma}^{\prime} \Big],	    \label{appRGvf}\\
			\frac{da}{dl}
			&=g^2 \Big( \frac{1 -a^2}{2a} 
			+ a (G_0 -G_1)  \Big) +a \sum_{\Gamma} \Delta_{\Gamma} +a \sum_{\Gamma} \Delta_{\Gamma}^{\prime},   \label{appRGa}\\
			\frac{dg^2}{dl}
			&=\epsilon g^2  -g^4 \big(1 -G_0 +3G_1\big) +\big( 4\Delta_S -\sum_{\Gamma} \Delta_{\Gamma} +3\sum_{\Gamma} {\Delta}_{\Gamma}^{\prime}\big) g^2 ,   \label{appRGg^2}\\
			\frac{d\Delta_S}{dl} 
			&= \big( \epsilon-1 \big) \Delta_S +2\Delta_S \big( \Delta_S+ \Delta_M +2 \Delta_V \big) +4\Delta_{M} \Delta_{V} +(G_0- 3G_2 -2G_1) \Delta_S g^2,    \label{appRGS}\\
			\frac{d\Delta_M}{dl} 
			&= 
			\big( \epsilon-1  \big) \Delta_M -2\Delta_{M} \big( \Delta_S+ \Delta_M -2 \Delta_V \big) +4\Delta_S \Delta_{V} -(G_0+ 3G_2+2G_1) \Delta_M g^2,    \label{appRGM}\\
			\frac{d\Delta_V}{dl}
			&= \big( \epsilon-1 \big) \Delta_V+
			2\Delta_{M} \Delta_S -(G_0+2G_1) \Delta_V g^2,     \label{appRGV}	\\
			\frac{d\Delta_S^{\prime}}{dl} 
			&= \big( \epsilon-1 \big) \Delta_S^{\prime} +2\Delta_S^{\prime} \big( \Delta_S+ \Delta_M +2 \Delta_V \big) +4\Delta_{M}^{\prime} \Delta_{V}^{\prime} +(G_0- 3G_2 -2G_1) \Delta_S^{\prime} g^2,    \label{appRGSp}\\
			\frac{d\Delta_M^{\prime}}{dl} 
			&= 
			\big( \epsilon-1  \big) \Delta_M^{\prime} -2\Delta_{M}^{\prime} \big( \Delta_S+ \Delta_M -2 \Delta_V \big) +4\Delta_S^{\prime} \Delta_{V}^{\prime} -(G_0+ 3G_2+2G_1) \Delta_M^{\prime} g^2,    \label{appRGMp}\\
			\frac{d\Delta_V^{\prime}}{dl}
			&= \big( \epsilon-1 \big) \Delta_V^{\prime}+
			2\Delta_{M}^{\prime} \Delta_S^{\prime} -(G_0+2G_1) \Delta_V^{\prime} g^2,     \label{appRGVp}
		\end{align}
	\end{widetext}
	Since the initial values of the parameters are $\Delta_{\Gamma}^{0}=\Delta_{\Gamma}^{\prime0}$, $\Delta_{\Gamma}$ and $\Delta_{\Gamma}^{\prime}$ will flow in the same way.
	We can simplify the equations by taken $\Delta_{\Gamma}^{\prime}=\Delta_{\Gamma}$ and produce the Eqs.~(\ref{RGvf})~-~(\ref{RGV}).
	
	For the case of $(3+1)D$, due to the change of disorder typies, the one-loop corrections of disorder couplings need to recalculate, the extension is direct for which we do not show details here.
\end{widetext}

\section*{Appendix C: general disorder case}
In this section, we study the case where all the disorder potential is appeared.
We focus on the regime near the Lorentz symmetric fixed point, $a=1+\delta a$ with small $\delta a\ll 1$.
The RG equations in 2D now become,
\begin{widetext}
	\begin{eqnarray}
		\frac{d v_f}{dl}
		&=&v_{f} \Big[ g^2 \frac{2}{3}\delta a -2\sum_{\Gamma} \Delta_{\mu} \Big],	    \\
		\frac{d\delta a}{dl}
		&=& \Big( -\frac{5}{3}g^2 +2\sum_{\Gamma} \Delta_{\Gamma} \Big) \delta a,  \\
		\frac{dg^2}{dl}
		&=&\epsilon g^2  -g^4 \big(3-2\delta a\big) +\big( 4\Delta_S +2\sum_{\Gamma} \Delta_{\Gamma} \big) g^2, \\
		\frac{d\Delta_S}{dl} 
		&=& \big( \epsilon-1 \big) \Delta_S +2\Delta_S \big( \Delta_S+ \Delta_M +2 \Delta_V \big) +4\Delta_{M} \Delta_{V} +(-4+\frac{17}{3}\delta a) \Delta_S g^2,  \\
		\frac{d\Delta_M}{dl} 
		&=& 
		\big( \epsilon-1  \big) \Delta_M -2\Delta_{M} \big( \Delta_S+ \Delta_M -2 \Delta_V \big) +4\Delta_S \Delta_{V} -(6 -\frac{29}{3}\delta a) \Delta_M g^2, \\
		\frac{d\Delta_V}{dl}
		&=& \big( \epsilon-1 \big) \Delta_V+
		2\Delta_{M} \Delta_S -(3-\frac{14}{3}\delta a) \Delta_V g^2,   
	\end{eqnarray}
\end{widetext}
There is a fixed point given by $(v_f, a^*,g^{2\ast}, \Delta_S^\ast, \Delta_M^\ast, \Delta_V^\ast) = (v_f, 1, \frac{\epsilon}3, 0 ,0, 0)$.
For small disorder case, the stability matrix at this fixed point is 
\begin{eqnarray}
	\left( \begin{array}{cccccc}
		0 & \frac29 \epsilon v_f & 0 & -2 v_f & -2 v_f & -2 v_f \\
		0 & -\frac59 \epsilon & 0 & 2 & 2 & 2 \\
		0 & \frac29 \epsilon^2 & -\epsilon & 2 \epsilon & \frac23 \epsilon & \frac23 \epsilon \\
		0 & 0 & 0 & -1- \frac{\epsilon}3 & 0 & 0 \\
		0 & 0 & 0 & 0 & -1-\epsilon & 0 \\
		0 & 0 & 0 & 0 & 0 & -1  
	\end{array}\right).
\end{eqnarray}
The eigenvalues of the stability matrix are all negative except for one marginal direction at $v_f$. 
We can conclude that the coupling between Yukawa potential and disorder will suppress weak random disorder potential. 
Similarly argument can also be applied to three dimensional case.

\bibliography{SUSY_dis_r2}

\begin{thebibliography}{84}%
\makeatletter
\providecommand \@ifxundefined [1]{%
 \@ifx{#1\undefined}
}%
\providecommand \@ifnum [1]{%
 \ifnum #1\expandafter \@firstoftwo
 \else \expandafter \@secondoftwo
 \fi
}%
\providecommand \@ifx [1]{%
 \ifx #1\expandafter \@firstoftwo
 \else \expandafter \@secondoftwo
 \fi
}%
\providecommand \natexlab [1]{#1}%
\providecommand \enquote  [1]{``#1''}%
\providecommand \bibnamefont  [1]{#1}%
\providecommand \bibfnamefont [1]{#1}%
\providecommand \citenamefont [1]{#1}%
\providecommand \href@noop [0]{\@secondoftwo}%
\providecommand \href [0]{\begingroup \@sanitize@url \@href}%
\providecommand \@href[1]{\@@startlink{#1}\@@href}%
\providecommand \@@href[1]{\endgroup#1\@@endlink}%
\providecommand \@sanitize@url [0]{\catcode `\\12\catcode `\$12\catcode
  `\&12\catcode `\#12\catcode `\^12\catcode `\_12\catcode `\%12\relax}%
\providecommand \@@startlink[1]{}%
\providecommand \@@endlink[0]{}%
\providecommand \url  [0]{\begingroup\@sanitize@url \@url }%
\providecommand \@url [1]{\endgroup\@href {#1}{\urlprefix }}%
\providecommand \urlprefix  [0]{URL }%
\providecommand \Eprint [0]{\href }%
\providecommand \doibase [0]{https://doi.org/}%
\providecommand \selectlanguage [0]{\@gobble}%
\providecommand \bibinfo  [0]{\@secondoftwo}%
\providecommand \bibfield  [0]{\@secondoftwo}%
\providecommand \translation [1]{[#1]}%
\providecommand \BibitemOpen [0]{}%
\providecommand \bibitemStop [0]{}%
\providecommand \bibitemNoStop [0]{.\EOS\space}%
\providecommand \EOS [0]{\spacefactor3000\relax}%
\providecommand \BibitemShut  [1]{\csname bibitem#1\endcsname}%
\let\auto@bib@innerbib\@empty
\bibitem [{\citenamefont {Weinberg}(2000)}]{Weinbergbook}%
  \BibitemOpen
  \bibfield  {author} {\bibinfo {author} {\bibfnamefont {S.}~\bibnamefont
  {Weinberg}},\ }\href
  {https://books.google.com.hk/books?id=QMkgAwAAQBAJ&printsec=frontcover&hl=zh-TW&source=gbs_ge_summary_r&cad=0#v=onepage&q&f=false}
  {\emph {\bibinfo {title} {The quantum theory of fields}}},\ Vol.~\bibinfo
  {volume} {3}\ (\bibinfo  {publisher} {Cambridge university press},\ \bibinfo
  {year} {2000})\BibitemShut {NoStop}%
\bibitem [{\citenamefont {Gervais}\ and\ \citenamefont
  {Sakita}(1971)}]{Gervais-Sakita1971}%
  \BibitemOpen
  \bibfield  {author} {\bibinfo {author} {\bibfnamefont {J.-L.}\ \bibnamefont
  {Gervais}}\ and\ \bibinfo {author} {\bibfnamefont {B.}~\bibnamefont
  {Sakita}},\ }\bibfield  {title} {\bibinfo {title} {Field theory
  interpretation of supergauges in dual models},\ }\href
  {https://doi.org/https://doi.org/10.1016/0550-3213(71)90351-8} {\bibfield
  {journal} {\bibinfo  {journal} {Nuclear Physics B}\ }\textbf {\bibinfo
  {volume} {34}},\ \bibinfo {pages} {632} (\bibinfo {year} {1971})}\BibitemShut
  {NoStop}%
\bibitem [{\citenamefont {Wess}\ and\ \citenamefont
  {Zumino}(1974)}]{Wess-Zumino1974}%
  \BibitemOpen
  \bibfield  {author} {\bibinfo {author} {\bibfnamefont {J.}~\bibnamefont
  {Wess}}\ and\ \bibinfo {author} {\bibfnamefont {B.}~\bibnamefont {Zumino}},\
  }\bibfield  {title} {\bibinfo {title} {Supergauge transformations in four
  dimensions},\ }\href
  {https://doi.org/https://doi.org/10.1016/0550-3213(74)90355-1} {\bibfield
  {journal} {\bibinfo  {journal} {Nuclear Physics B}\ }\textbf {\bibinfo
  {volume} {70}},\ \bibinfo {pages} {39} (\bibinfo {year} {1974})}\BibitemShut
  {NoStop}%
\bibitem [{\citenamefont {Dimopoulos}\ and\ \citenamefont
  {Georgi}(1981)}]{Dimopoulos-Georgi1981}%
  \BibitemOpen
  \bibfield  {author} {\bibinfo {author} {\bibfnamefont {S.}~\bibnamefont
  {Dimopoulos}}\ and\ \bibinfo {author} {\bibfnamefont {H.}~\bibnamefont
  {Georgi}},\ }\bibfield  {title} {\bibinfo {title} {Softly broken
  supersymmetry and su(5)},\ }\href
  {https://doi.org/https://doi.org/10.1016/0550-3213(81)90522-8} {\bibfield
  {journal} {\bibinfo  {journal} {Nuclear Physics B}\ }\textbf {\bibinfo
  {volume} {193}},\ \bibinfo {pages} {150} (\bibinfo {year}
  {1981})}\BibitemShut {NoStop}%
\bibitem [{\citenamefont {Cremmer}\ \emph {et~al.}(1983)\citenamefont
  {Cremmer}, \citenamefont {Ferrara}, \citenamefont {Kounnas},\ and\
  \citenamefont {Nanopoulos}}]{Cremmer1983}%
  \BibitemOpen
  \bibfield  {author} {\bibinfo {author} {\bibfnamefont {E.}~\bibnamefont
  {Cremmer}}, \bibinfo {author} {\bibfnamefont {S.}~\bibnamefont {Ferrara}},
  \bibinfo {author} {\bibfnamefont {C.}~\bibnamefont {Kounnas}},\ and\ \bibinfo
  {author} {\bibfnamefont {D.}~\bibnamefont {Nanopoulos}},\ }\bibfield  {title}
  {\bibinfo {title} {Naturally vanishing cosmological constant in n=1
  supergravity},\ }\href
  {https://doi.org/https://doi.org/10.1016/0370-2693(83)90106-5} {\bibfield
  {journal} {\bibinfo  {journal} {Physics Letters B}\ }\textbf {\bibinfo
  {volume} {133}},\ \bibinfo {pages} {61} (\bibinfo {year} {1983})}\BibitemShut
  {NoStop}%
\bibitem [{\citenamefont {Seiberg}\ and\ \citenamefont {Witten}(1994)}]{SW}%
  \BibitemOpen
  \bibfield  {author} {\bibinfo {author} {\bibfnamefont {N.}~\bibnamefont
  {Seiberg}}\ and\ \bibinfo {author} {\bibfnamefont {E.}~\bibnamefont
  {Witten}},\ }\bibfield  {title} {\bibinfo {title} {Electric-magnetic duality,
  monopole condensation, and confinement in n=2 supersymmetric yang-mills
  theory},\ }\href
  {https://doi.org/https://doi.org/10.1016/0550-3213(94)90124-4} {\bibfield
  {journal} {\bibinfo  {journal} {Nuclear Physics B}\ }\textbf {\bibinfo
  {volume} {426}},\ \bibinfo {pages} {19} (\bibinfo {year} {1994})}\BibitemShut
  {NoStop}%
\bibitem [{\citenamefont {Seiberg}(1995)}]{SEIBERG}%
  \BibitemOpen
  \bibfield  {author} {\bibinfo {author} {\bibfnamefont {N.}~\bibnamefont
  {Seiberg}},\ }\bibfield  {title} {\bibinfo {title} {Electric-magnetic duality
  in supersymmetric non-abelian gauge theories},\ }\href
  {https://doi.org/https://doi.org/10.1016/0550-3213(94)00023-8} {\bibfield
  {journal} {\bibinfo  {journal} {Nuclear Physics B}\ }\textbf {\bibinfo
  {volume} {435}},\ \bibinfo {pages} {129} (\bibinfo {year}
  {1995})}\BibitemShut {NoStop}%
\bibitem [{\citenamefont {Wan}\ \emph {et~al.}(2011)\citenamefont {Wan},
  \citenamefont {Turner}, \citenamefont {Vishwanath},\ and\ \citenamefont
  {Savrasov}}]{ashvin2011}%
  \BibitemOpen
  \bibfield  {author} {\bibinfo {author} {\bibfnamefont {X.}~\bibnamefont
  {Wan}}, \bibinfo {author} {\bibfnamefont {A.~M.}\ \bibnamefont {Turner}},
  \bibinfo {author} {\bibfnamefont {A.}~\bibnamefont {Vishwanath}},\ and\
  \bibinfo {author} {\bibfnamefont {S.~Y.}\ \bibnamefont {Savrasov}},\
  }\bibfield  {title} {\bibinfo {title} {Topological semimetal and fermi-arc
  surface states in the electronic structure of pyrochlore iridates},\ }\href
  {https://doi.org/10.1103/PhysRevB.83.205101} {\bibfield  {journal} {\bibinfo
  {journal} {Phys. Rev. B}\ }\textbf {\bibinfo {volume} {83}},\ \bibinfo
  {pages} {205101} (\bibinfo {year} {2011})}\BibitemShut {NoStop}%
\bibitem [{\citenamefont {Xu}\ \emph {et~al.}(2011)\citenamefont {Xu},
  \citenamefont {Weng}, \citenamefont {Wang}, \citenamefont {Dai},\ and\
  \citenamefont {Fang}}]{xu2011}%
  \BibitemOpen
  \bibfield  {author} {\bibinfo {author} {\bibfnamefont {G.}~\bibnamefont
  {Xu}}, \bibinfo {author} {\bibfnamefont {H.}~\bibnamefont {Weng}}, \bibinfo
  {author} {\bibfnamefont {Z.}~\bibnamefont {Wang}}, \bibinfo {author}
  {\bibfnamefont {X.}~\bibnamefont {Dai}},\ and\ \bibinfo {author}
  {\bibfnamefont {Z.}~\bibnamefont {Fang}},\ }\bibfield  {title} {\bibinfo
  {title} {Chern semimetal and the quantized anomalous hall effect in
  ${\mathrm{hgcr}}_{2}{\mathrm{se}}_{4}$},\ }\href
  {https://doi.org/10.1103/PhysRevLett.107.186806} {\bibfield  {journal}
  {\bibinfo  {journal} {Phys. Rev. Lett.}\ }\textbf {\bibinfo {volume} {107}},\
  \bibinfo {pages} {186806} (\bibinfo {year} {2011})}\BibitemShut {NoStop}%
\bibitem [{\citenamefont {Burkov}\ and\ \citenamefont
  {Balents}(2011)}]{burkov2011}%
  \BibitemOpen
  \bibfield  {author} {\bibinfo {author} {\bibfnamefont {A.~A.}\ \bibnamefont
  {Burkov}}\ and\ \bibinfo {author} {\bibfnamefont {L.}~\bibnamefont
  {Balents}},\ }\bibfield  {title} {\bibinfo {title} {Weyl semimetal in a
  topological insulator multilayer},\ }\href
  {https://doi.org/10.1103/PhysRevLett.107.127205} {\bibfield  {journal}
  {\bibinfo  {journal} {Phys. Rev. Lett.}\ }\textbf {\bibinfo {volume} {107}},\
  \bibinfo {pages} {127205} (\bibinfo {year} {2011})}\BibitemShut {NoStop}%
\bibitem [{\citenamefont {Weng}\ \emph {et~al.}(2015)\citenamefont {Weng},
  \citenamefont {Fang}, \citenamefont {Fang}, \citenamefont {Bernevig},\ and\
  \citenamefont {Dai}}]{weng2015}%
  \BibitemOpen
  \bibfield  {author} {\bibinfo {author} {\bibfnamefont {H.}~\bibnamefont
  {Weng}}, \bibinfo {author} {\bibfnamefont {C.}~\bibnamefont {Fang}}, \bibinfo
  {author} {\bibfnamefont {Z.}~\bibnamefont {Fang}}, \bibinfo {author}
  {\bibfnamefont {B.~A.}\ \bibnamefont {Bernevig}},\ and\ \bibinfo {author}
  {\bibfnamefont {X.}~\bibnamefont {Dai}},\ }\bibfield  {title} {\bibinfo
  {title} {Weyl semimetal phase in noncentrosymmetric transition-metal
  monophosphides},\ }\href {https://doi.org/10.1103/PhysRevX.5.011029}
  {\bibfield  {journal} {\bibinfo  {journal} {Phys. Rev. X}\ }\textbf {\bibinfo
  {volume} {5}},\ \bibinfo {pages} {011029} (\bibinfo {year}
  {2015})}\BibitemShut {NoStop}%
\bibitem [{\citenamefont {Huang}\ \emph {et~al.}(2015)\citenamefont {Huang},
  \citenamefont {Xu}, \citenamefont {Belopolski}, \citenamefont {Lee},
  \citenamefont {Chang}, \citenamefont {Wang}, \citenamefont {Alidoust},
  \citenamefont {Bian}, \citenamefont {Neupane}, \citenamefont {Zhang} \emph
  {et~al.}}]{hasan2015}%
  \BibitemOpen
  \bibfield  {author} {\bibinfo {author} {\bibfnamefont {S.-M.}\ \bibnamefont
  {Huang}}, \bibinfo {author} {\bibfnamefont {S.-Y.}\ \bibnamefont {Xu}},
  \bibinfo {author} {\bibfnamefont {I.}~\bibnamefont {Belopolski}}, \bibinfo
  {author} {\bibfnamefont {C.-C.}\ \bibnamefont {Lee}}, \bibinfo {author}
  {\bibfnamefont {G.}~\bibnamefont {Chang}}, \bibinfo {author} {\bibfnamefont
  {B.}~\bibnamefont {Wang}}, \bibinfo {author} {\bibfnamefont {N.}~\bibnamefont
  {Alidoust}}, \bibinfo {author} {\bibfnamefont {G.}~\bibnamefont {Bian}},
  \bibinfo {author} {\bibfnamefont {M.}~\bibnamefont {Neupane}}, \bibinfo
  {author} {\bibfnamefont {C.}~\bibnamefont {Zhang}}, \emph {et~al.},\
  }\bibfield  {title} {\bibinfo {title} {A weyl fermion semimetal with surface
  fermi arcs in the transition metal monopnictide taas class},\ }\href
  {https://www.nature.com/articles/ncomms8373#citeas} {\bibfield  {journal}
  {\bibinfo  {journal} {Nature communications}\ }\textbf {\bibinfo {volume}
  {6}},\ \bibinfo {pages} {1} (\bibinfo {year} {2015})}\BibitemShut {NoStop}%
\bibitem [{\citenamefont {Xu}\ \emph {et~al.}(2015)\citenamefont {Xu},
  \citenamefont {Belopolski}, \citenamefont {Alidoust}, \citenamefont
  {Neupane}, \citenamefont {Bian}, \citenamefont {Zhang}, \citenamefont
  {Sankar}, \citenamefont {Chang}, \citenamefont {Yuan}, \citenamefont {Lee}
  \emph {et~al.}}]{xu2015}%
  \BibitemOpen
  \bibfield  {author} {\bibinfo {author} {\bibfnamefont {S.-Y.}\ \bibnamefont
  {Xu}}, \bibinfo {author} {\bibfnamefont {I.}~\bibnamefont {Belopolski}},
  \bibinfo {author} {\bibfnamefont {N.}~\bibnamefont {Alidoust}}, \bibinfo
  {author} {\bibfnamefont {M.}~\bibnamefont {Neupane}}, \bibinfo {author}
  {\bibfnamefont {G.}~\bibnamefont {Bian}}, \bibinfo {author} {\bibfnamefont
  {C.}~\bibnamefont {Zhang}}, \bibinfo {author} {\bibfnamefont
  {R.}~\bibnamefont {Sankar}}, \bibinfo {author} {\bibfnamefont
  {G.}~\bibnamefont {Chang}}, \bibinfo {author} {\bibfnamefont
  {Z.}~\bibnamefont {Yuan}}, \bibinfo {author} {\bibfnamefont {C.-C.}\
  \bibnamefont {Lee}}, \emph {et~al.},\ }\bibfield  {title} {\bibinfo {title}
  {Discovery of a weyl fermion semimetal and topological fermi arcs},\ }\href
  {https://www.science.org/doi/full/10.1126/science.aaa9297} {\bibfield
  {journal} {\bibinfo  {journal} {Science}\ }\textbf {\bibinfo {volume}
  {349}},\ \bibinfo {pages} {613} (\bibinfo {year} {2015})}\BibitemShut
  {NoStop}%
\bibitem [{\citenamefont {Lv}\ \emph {et~al.}(2015)\citenamefont {Lv},
  \citenamefont {Weng}, \citenamefont {Fu}, \citenamefont {Wang}, \citenamefont
  {Miao}, \citenamefont {Ma}, \citenamefont {Richard}, \citenamefont {Huang},
  \citenamefont {Zhao}, \citenamefont {Chen}, \citenamefont {Fang},
  \citenamefont {Dai}, \citenamefont {Qian},\ and\ \citenamefont
  {Ding}}]{lv2015}%
  \BibitemOpen
  \bibfield  {author} {\bibinfo {author} {\bibfnamefont {B.~Q.}\ \bibnamefont
  {Lv}}, \bibinfo {author} {\bibfnamefont {H.~M.}\ \bibnamefont {Weng}},
  \bibinfo {author} {\bibfnamefont {B.~B.}\ \bibnamefont {Fu}}, \bibinfo
  {author} {\bibfnamefont {X.~P.}\ \bibnamefont {Wang}}, \bibinfo {author}
  {\bibfnamefont {H.}~\bibnamefont {Miao}}, \bibinfo {author} {\bibfnamefont
  {J.}~\bibnamefont {Ma}}, \bibinfo {author} {\bibfnamefont {P.}~\bibnamefont
  {Richard}}, \bibinfo {author} {\bibfnamefont {X.~C.}\ \bibnamefont {Huang}},
  \bibinfo {author} {\bibfnamefont {L.~X.}\ \bibnamefont {Zhao}}, \bibinfo
  {author} {\bibfnamefont {G.~F.}\ \bibnamefont {Chen}}, \bibinfo {author}
  {\bibfnamefont {Z.}~\bibnamefont {Fang}}, \bibinfo {author} {\bibfnamefont
  {X.}~\bibnamefont {Dai}}, \bibinfo {author} {\bibfnamefont {T.}~\bibnamefont
  {Qian}},\ and\ \bibinfo {author} {\bibfnamefont {H.}~\bibnamefont {Ding}},\
  }\bibfield  {title} {\bibinfo {title} {Experimental discovery of weyl
  semimetal taas},\ }\href {https://doi.org/10.1103/PhysRevX.5.031013}
  {\bibfield  {journal} {\bibinfo  {journal} {Phys. Rev. X}\ }\textbf {\bibinfo
  {volume} {5}},\ \bibinfo {pages} {031013} (\bibinfo {year}
  {2015})}\BibitemShut {NoStop}%
\bibitem [{\citenamefont {Yang}\ \emph {et~al.}(2015)\citenamefont {Yang},
  \citenamefont {Liu}, \citenamefont {Sun}, \citenamefont {Peng}, \citenamefont
  {Yang}, \citenamefont {Zhang}, \citenamefont {Zhou}, \citenamefont {Zhang},
  \citenamefont {Guo}, \citenamefont {Rahn} \emph {et~al.}}]{yang2015}%
  \BibitemOpen
  \bibfield  {author} {\bibinfo {author} {\bibfnamefont {L.}~\bibnamefont
  {Yang}}, \bibinfo {author} {\bibfnamefont {Z.}~\bibnamefont {Liu}}, \bibinfo
  {author} {\bibfnamefont {Y.}~\bibnamefont {Sun}}, \bibinfo {author}
  {\bibfnamefont {H.}~\bibnamefont {Peng}}, \bibinfo {author} {\bibfnamefont
  {H.}~\bibnamefont {Yang}}, \bibinfo {author} {\bibfnamefont {T.}~\bibnamefont
  {Zhang}}, \bibinfo {author} {\bibfnamefont {B.}~\bibnamefont {Zhou}},
  \bibinfo {author} {\bibfnamefont {Y.}~\bibnamefont {Zhang}}, \bibinfo
  {author} {\bibfnamefont {Y.}~\bibnamefont {Guo}}, \bibinfo {author}
  {\bibfnamefont {M.}~\bibnamefont {Rahn}}, \emph {et~al.},\ }\bibfield
  {title} {\bibinfo {title} {Weyl semimetal phase in the non-centrosymmetric
  compound taas},\ }\href {https://www.nature.com/articles/nphys3425}
  {\bibfield  {journal} {\bibinfo  {journal} {Nature physics}\ }\textbf
  {\bibinfo {volume} {11}},\ \bibinfo {pages} {728} (\bibinfo {year}
  {2015})}\BibitemShut {NoStop}%
\bibitem [{\citenamefont {Curci}\ and\ \citenamefont
  {Veneziano}(1987)}]{CURCI}%
  \BibitemOpen
  \bibfield  {author} {\bibinfo {author} {\bibfnamefont {G.}~\bibnamefont
  {Curci}}\ and\ \bibinfo {author} {\bibfnamefont {G.}~\bibnamefont
  {Veneziano}},\ }\bibfield  {title} {\bibinfo {title} {Supersymmetry and the
  lattice: A reconciliation?},\ }\href
  {https://doi.org/https://doi.org/10.1016/0550-3213(87)90660-2} {\bibfield
  {journal} {\bibinfo  {journal} {Nuclear Physics B}\ }\textbf {\bibinfo
  {volume} {292}},\ \bibinfo {pages} {555} (\bibinfo {year}
  {1987})}\BibitemShut {NoStop}%
\bibitem [{\citenamefont {Goh}\ \emph {et~al.}(2005)\citenamefont {Goh},
  \citenamefont {Luty},\ and\ \citenamefont {Ng}}]{GOH}%
  \BibitemOpen
  \bibfield  {author} {\bibinfo {author} {\bibfnamefont {H.-S.}\ \bibnamefont
  {Goh}}, \bibinfo {author} {\bibfnamefont {M.~A.}\ \bibnamefont {Luty}},\ and\
  \bibinfo {author} {\bibfnamefont {S.-P.}\ \bibnamefont {Ng}},\ }\bibfield
  {title} {\bibinfo {title} {Supersymmetry without supersymmetry},\ }\href
  {https://doi.org/10.1088/1126-6708/2005/01/040} {\bibfield  {journal}
  {\bibinfo  {journal} {Journal of High Energy Physics}\ }\textbf {\bibinfo
  {volume} {2005}},\ \bibinfo {pages} {040} (\bibinfo {year}
  {2005})}\BibitemShut {NoStop}%
\bibitem [{\citenamefont {Thomas}()}]{SCOTT}%
  \BibitemOpen
  \bibfield  {author} {\bibinfo {author} {\bibfnamefont {S.}~\bibnamefont
  {Thomas}},\ }\href@noop {} {\bibinfo {title} {Emergent supersymmetry, kitp
  talk, 2005}}\BibitemShut {NoStop}%
\bibitem [{\citenamefont {Prakash}\ and\ \citenamefont
  {Wang}(2021{\natexlab{a}})}]{juven2021}%
  \BibitemOpen
  \bibfield  {author} {\bibinfo {author} {\bibfnamefont {A.}~\bibnamefont
  {Prakash}}\ and\ \bibinfo {author} {\bibfnamefont {J.}~\bibnamefont {Wang}},\
  }\bibfield  {title} {\bibinfo {title} {Boundary supersymmetry of
  $(1+1)\mathrm{D}$ fermionic symmetry-protected topological phases},\ }\href
  {https://doi.org/10.1103/PhysRevLett.126.236802} {\bibfield  {journal}
  {\bibinfo  {journal} {Phys. Rev. Lett.}\ }\textbf {\bibinfo {volume} {126}},\
  \bibinfo {pages} {236802} (\bibinfo {year} {2021}{\natexlab{a}})}\BibitemShut
  {NoStop}%
\bibitem [{\citenamefont {Prakash}\ and\ \citenamefont
  {Wang}(2021{\natexlab{b}})}]{juven2021prb}%
  \BibitemOpen
  \bibfield  {author} {\bibinfo {author} {\bibfnamefont {A.}~\bibnamefont
  {Prakash}}\ and\ \bibinfo {author} {\bibfnamefont {J.}~\bibnamefont {Wang}},\
  }\bibfield  {title} {\bibinfo {title} {Unwinding fermionic symmetry-protected
  topological phases: Supersymmetry extension},\ }\href
  {https://doi.org/10.1103/PhysRevB.103.085130} {\bibfield  {journal} {\bibinfo
   {journal} {Phys. Rev. B}\ }\textbf {\bibinfo {volume} {103}},\ \bibinfo
  {pages} {085130} (\bibinfo {year} {2021}{\natexlab{b}})}\BibitemShut
  {NoStop}%
\bibitem [{\citenamefont {Turzillo}\ and\ \citenamefont
  {You}(2021)}]{Alex2021}%
  \BibitemOpen
  \bibfield  {author} {\bibinfo {author} {\bibfnamefont {A.}~\bibnamefont
  {Turzillo}}\ and\ \bibinfo {author} {\bibfnamefont {M.}~\bibnamefont {You}},\
  }\bibfield  {title} {\bibinfo {title} {Supersymmetric boundaries of
  one-dimensional phases of fermions beyond symmetry-protected topological
  states},\ }\href {https://doi.org/10.1103/PhysRevLett.127.026402} {\bibfield
  {journal} {\bibinfo  {journal} {Phys. Rev. Lett.}\ }\textbf {\bibinfo
  {volume} {127}},\ \bibinfo {pages} {026402} (\bibinfo {year}
  {2021})}\BibitemShut {NoStop}%
\bibitem [{\citenamefont {Li}\ \emph {et~al.}(2018)\citenamefont {Li},
  \citenamefont {Vaezi}, \citenamefont {Mendl},\ and\ \citenamefont
  {Yao}}]{zixiang2018sa}%
  \BibitemOpen
  \bibfield  {author} {\bibinfo {author} {\bibfnamefont {Z.-X.}\ \bibnamefont
  {Li}}, \bibinfo {author} {\bibfnamefont {A.}~\bibnamefont {Vaezi}}, \bibinfo
  {author} {\bibfnamefont {C.~B.}\ \bibnamefont {Mendl}},\ and\ \bibinfo
  {author} {\bibfnamefont {H.}~\bibnamefont {Yao}},\ }\bibfield  {title}
  {\bibinfo {title} {Numerical observation of emergent spacetime supersymmetry
  at quantum criticality},\ }\href
  {https://www.science.org/doi/full/10.1126/sciadv.aau1463} {\bibfield
  {journal} {\bibinfo  {journal} {Science advances}\ }\textbf {\bibinfo
  {volume} {4}},\ \bibinfo {pages} {eaau1463} (\bibinfo {year}
  {2018})}\BibitemShut {NoStop}%
\bibitem [{\citenamefont {Li}\ and\ \citenamefont
  {Yao}(2017)}]{zixiang2017prb}%
  \BibitemOpen
  \bibfield  {author} {\bibinfo {author} {\bibfnamefont {Z.-X.}\ \bibnamefont
  {Li}}\ and\ \bibinfo {author} {\bibfnamefont {H.}~\bibnamefont {Yao}},\
  }\bibfield  {title} {\bibinfo {title} {Edge stability and edge quantum
  criticality in two-dimensional interacting topological insulators},\ }\href
  {https://doi.org/10.1103/PhysRevB.96.241101} {\bibfield  {journal} {\bibinfo
  {journal} {Phys. Rev. B}\ }\textbf {\bibinfo {volume} {96}},\ \bibinfo
  {pages} {241101} (\bibinfo {year} {2017})}\BibitemShut {NoStop}%
\bibitem [{\citenamefont {Li}\ \emph {et~al.}(2016)\citenamefont {Li},
  \citenamefont {Jiang},\ and\ \citenamefont {Yao}}]{zixiang2017prl}%
  \BibitemOpen
  \bibfield  {author} {\bibinfo {author} {\bibfnamefont {Z.-X.}\ \bibnamefont
  {Li}}, \bibinfo {author} {\bibfnamefont {Y.-F.}\ \bibnamefont {Jiang}},\ and\
  \bibinfo {author} {\bibfnamefont {H.}~\bibnamefont {Yao}},\ }\bibfield
  {title} {\bibinfo {title} {Majorana-time-reversal symmetries: A fundamental
  principle for sign-problem-free quantum monte carlo simulations},\ }\href
  {https://doi.org/10.1103/PhysRevLett.117.267002} {\bibfield  {journal}
  {\bibinfo  {journal} {Phys. Rev. Lett.}\ }\textbf {\bibinfo {volume} {117}},\
  \bibinfo {pages} {267002} (\bibinfo {year} {2016})}\BibitemShut {NoStop}%
\bibitem [{\citenamefont {Jian}\ \emph {et~al.}(2017)\citenamefont {Jian},
  \citenamefont {Lin}, \citenamefont {Maciejko},\ and\ \citenamefont
  {Yao}}]{shaokai2017prl}%
  \BibitemOpen
  \bibfield  {author} {\bibinfo {author} {\bibfnamefont {S.-K.}\ \bibnamefont
  {Jian}}, \bibinfo {author} {\bibfnamefont {C.-H.}\ \bibnamefont {Lin}},
  \bibinfo {author} {\bibfnamefont {J.}~\bibnamefont {Maciejko}},\ and\
  \bibinfo {author} {\bibfnamefont {H.}~\bibnamefont {Yao}},\ }\bibfield
  {title} {\bibinfo {title} {Emergence of supersymmetric quantum
  electrodynamics},\ }\href {https://doi.org/10.1103/PhysRevLett.118.166802}
  {\bibfield  {journal} {\bibinfo  {journal} {Phys. Rev. Lett.}\ }\textbf
  {\bibinfo {volume} {118}},\ \bibinfo {pages} {166802} (\bibinfo {year}
  {2017})}\BibitemShut {NoStop}%
\bibitem [{\citenamefont {Yu}\ \emph {et~al.}(2019)\citenamefont {Yu},
  \citenamefont {Roiban}, \citenamefont {Jian},\ and\ \citenamefont
  {Liu}}]{Yu2019prb}%
  \BibitemOpen
  \bibfield  {author} {\bibinfo {author} {\bibfnamefont {J.}~\bibnamefont
  {Yu}}, \bibinfo {author} {\bibfnamefont {R.}~\bibnamefont {Roiban}}, \bibinfo
  {author} {\bibfnamefont {S.-K.}\ \bibnamefont {Jian}},\ and\ \bibinfo
  {author} {\bibfnamefont {C.-X.}\ \bibnamefont {Liu}},\ }\bibfield  {title}
  {\bibinfo {title} {Finite-scale emergence of $2+1\mathrm{D}$ supersymmetry at
  first-order quantum phase transition},\ }\href
  {https://doi.org/10.1103/PhysRevB.100.075153} {\bibfield  {journal} {\bibinfo
   {journal} {Phys. Rev. B}\ }\textbf {\bibinfo {volume} {100}},\ \bibinfo
  {pages} {075153} (\bibinfo {year} {2019})}\BibitemShut {NoStop}%
\bibitem [{\citenamefont {Lee}(2007)}]{lee2007}%
  \BibitemOpen
  \bibfield  {author} {\bibinfo {author} {\bibfnamefont {S.-S.}\ \bibnamefont
  {Lee}},\ }\bibfield  {title} {\bibinfo {title} {Emergence of supersymmetry at
  a critical point of a lattice model},\ }\href
  {https://doi.org/10.1103/PhysRevB.76.075103} {\bibfield  {journal} {\bibinfo
  {journal} {Phys. Rev. B}\ }\textbf {\bibinfo {volume} {76}},\ \bibinfo
  {pages} {075103} (\bibinfo {year} {2007})}\BibitemShut {NoStop}%
\bibitem [{\citenamefont {Yu}\ and\ \citenamefont {Yang}(2010)}]{yu2010}%
  \BibitemOpen
  \bibfield  {author} {\bibinfo {author} {\bibfnamefont {Y.}~\bibnamefont
  {Yu}}\ and\ \bibinfo {author} {\bibfnamefont {K.}~\bibnamefont {Yang}},\
  }\bibfield  {title} {\bibinfo {title} {Simulating the wess-zumino
  supersymmetry model in optical lattices},\ }\href
  {https://doi.org/10.1103/PhysRevLett.105.150605} {\bibfield  {journal}
  {\bibinfo  {journal} {Phys. Rev. Lett.}\ }\textbf {\bibinfo {volume} {105}},\
  \bibinfo {pages} {150605} (\bibinfo {year} {2010})}\BibitemShut {NoStop}%
\bibitem [{\citenamefont {Hasan}\ and\ \citenamefont {Kane}(2010)}]{kane2010}%
  \BibitemOpen
  \bibfield  {author} {\bibinfo {author} {\bibfnamefont {M.~Z.}\ \bibnamefont
  {Hasan}}\ and\ \bibinfo {author} {\bibfnamefont {C.~L.}\ \bibnamefont
  {Kane}},\ }\bibfield  {title} {\bibinfo {title} {Colloquium: Topological
  insulators},\ }\href {https://doi.org/10.1103/RevModPhys.82.3045} {\bibfield
  {journal} {\bibinfo  {journal} {Rev. Mod. Phys.}\ }\textbf {\bibinfo {volume}
  {82}},\ \bibinfo {pages} {3045} (\bibinfo {year} {2010})}\BibitemShut
  {NoStop}%
\bibitem [{\citenamefont {Qi}\ and\ \citenamefont {Zhang}(2011)}]{qi2011}%
  \BibitemOpen
  \bibfield  {author} {\bibinfo {author} {\bibfnamefont {X.-L.}\ \bibnamefont
  {Qi}}\ and\ \bibinfo {author} {\bibfnamefont {S.-C.}\ \bibnamefont {Zhang}},\
  }\bibfield  {title} {\bibinfo {title} {Topological insulators and
  superconductors},\ }\href {https://doi.org/10.1103/RevModPhys.83.1057}
  {\bibfield  {journal} {\bibinfo  {journal} {Rev. Mod. Phys.}\ }\textbf
  {\bibinfo {volume} {83}},\ \bibinfo {pages} {1057} (\bibinfo {year}
  {2011})}\BibitemShut {NoStop}%
\bibitem [{\citenamefont {Grover}\ \emph {et~al.}(2014)\citenamefont {Grover},
  \citenamefont {Sheng},\ and\ \citenamefont {Vishwanath}}]{grover2014}%
  \BibitemOpen
  \bibfield  {author} {\bibinfo {author} {\bibfnamefont {T.}~\bibnamefont
  {Grover}}, \bibinfo {author} {\bibfnamefont {D.}~\bibnamefont {Sheng}},\ and\
  \bibinfo {author} {\bibfnamefont {A.}~\bibnamefont {Vishwanath}},\ }\bibfield
   {title} {\bibinfo {title} {Emergent space-time supersymmetry at the boundary
  of a topological phase},\ }\href
  {https://www.science.org/doi/full/10.1126/science.1248253?casa_token=zapkame37VkAAAAA%3A3EGx393WteQ9Egv9BYYSnk9Xqid3KfogR4gp82J9XpHJ4qwJPc20WwQG0LJR2ydSn9RbLs8nd9k2bg}
  {\bibfield  {journal} {\bibinfo  {journal} {Science}\ }\textbf {\bibinfo
  {volume} {344}},\ \bibinfo {pages} {280} (\bibinfo {year}
  {2014})}\BibitemShut {NoStop}%
\bibitem [{\citenamefont {Ponte}\ and\ \citenamefont {Lee}(2014)}]{ponte2014}%
  \BibitemOpen
  \bibfield  {author} {\bibinfo {author} {\bibfnamefont {P.}~\bibnamefont
  {Ponte}}\ and\ \bibinfo {author} {\bibfnamefont {S.-S.}\ \bibnamefont
  {Lee}},\ }\bibfield  {title} {\bibinfo {title} {Emergence of supersymmetry on
  the surface of three-dimensional topological insulators},\ }\href
  {https://doi.org/10.1088/1367-2630/16/1/013044} {\bibfield  {journal}
  {\bibinfo  {journal} {New Journal of Physics}\ }\textbf {\bibinfo {volume}
  {16}},\ \bibinfo {pages} {013044} (\bibinfo {year} {2014})}\BibitemShut
  {NoStop}%
\bibitem [{\citenamefont {Zerf}\ \emph {et~al.}(2016)\citenamefont {Zerf},
  \citenamefont {Lin},\ and\ \citenamefont {Maciejko}}]{zerf2016}%
  \BibitemOpen
  \bibfield  {author} {\bibinfo {author} {\bibfnamefont {N.}~\bibnamefont
  {Zerf}}, \bibinfo {author} {\bibfnamefont {C.-H.}\ \bibnamefont {Lin}},\ and\
  \bibinfo {author} {\bibfnamefont {J.}~\bibnamefont {Maciejko}},\ }\bibfield
  {title} {\bibinfo {title} {Superconducting quantum criticality of topological
  surface states at three loops},\ }\href
  {https://doi.org/10.1103/PhysRevB.94.205106} {\bibfield  {journal} {\bibinfo
  {journal} {Phys. Rev. B}\ }\textbf {\bibinfo {volume} {94}},\ \bibinfo
  {pages} {205106} (\bibinfo {year} {2016})}\BibitemShut {NoStop}%
\bibitem [{\citenamefont {Friedan}\ \emph {et~al.}(1984)\citenamefont
  {Friedan}, \citenamefont {Qiu},\ and\ \citenamefont {Shenker}}]{friedan1984}%
  \BibitemOpen
  \bibfield  {author} {\bibinfo {author} {\bibfnamefont {D.}~\bibnamefont
  {Friedan}}, \bibinfo {author} {\bibfnamefont {Z.}~\bibnamefont {Qiu}},\ and\
  \bibinfo {author} {\bibfnamefont {S.}~\bibnamefont {Shenker}},\ }\bibfield
  {title} {\bibinfo {title} {Conformal invariance, unitarity, and critical
  exponents in two dimensions},\ }\href
  {https://doi.org/10.1103/PhysRevLett.52.1575} {\bibfield  {journal} {\bibinfo
   {journal} {Phys. Rev. Lett.}\ }\textbf {\bibinfo {volume} {52}},\ \bibinfo
  {pages} {1575} (\bibinfo {year} {1984})}\BibitemShut {NoStop}%
\bibitem [{\citenamefont {Foda}(1988)}]{foda1988}%
  \BibitemOpen
  \bibfield  {author} {\bibinfo {author} {\bibfnamefont {O.}~\bibnamefont
  {Foda}},\ }\bibfield  {title} {\bibinfo {title} {A supersymmetric phase
  transition in josephson-tunnel-junction arrays},\ }\href
  {https://doi.org/https://doi.org/10.1016/0550-3213(88)90615-3} {\bibfield
  {journal} {\bibinfo  {journal} {Nuclear Physics B}\ }\textbf {\bibinfo
  {volume} {300}},\ \bibinfo {pages} {611} (\bibinfo {year}
  {1988})}\BibitemShut {NoStop}%
\bibitem [{\citenamefont {Huijse}\ \emph {et~al.}(2015)\citenamefont {Huijse},
  \citenamefont {Bauer},\ and\ \citenamefont {Berg}}]{huijse2015}%
  \BibitemOpen
  \bibfield  {author} {\bibinfo {author} {\bibfnamefont {L.}~\bibnamefont
  {Huijse}}, \bibinfo {author} {\bibfnamefont {B.}~\bibnamefont {Bauer}},\ and\
  \bibinfo {author} {\bibfnamefont {E.}~\bibnamefont {Berg}},\ }\bibfield
  {title} {\bibinfo {title} {Emergent supersymmetry at the
  ising--berezinskii-kosterlitz-thouless multicritical point},\ }\href
  {https://doi.org/10.1103/PhysRevLett.114.090404} {\bibfield  {journal}
  {\bibinfo  {journal} {Phys. Rev. Lett.}\ }\textbf {\bibinfo {volume} {114}},\
  \bibinfo {pages} {090404} (\bibinfo {year} {2015})}\BibitemShut {NoStop}%
\bibitem [{\citenamefont {Jian}\ \emph {et~al.}(2015)\citenamefont {Jian},
  \citenamefont {Jiang},\ and\ \citenamefont {Yao}}]{jian2015}%
  \BibitemOpen
  \bibfield  {author} {\bibinfo {author} {\bibfnamefont {S.-K.}\ \bibnamefont
  {Jian}}, \bibinfo {author} {\bibfnamefont {Y.-F.}\ \bibnamefont {Jiang}},\
  and\ \bibinfo {author} {\bibfnamefont {H.}~\bibnamefont {Yao}},\ }\bibfield
  {title} {\bibinfo {title} {Emergent spacetime supersymmetry in 3d weyl
  semimetals and 2d dirac semimetals},\ }\href
  {https://doi.org/10.1103/PhysRevLett.114.237001} {\bibfield  {journal}
  {\bibinfo  {journal} {Phys. Rev. Lett.}\ }\textbf {\bibinfo {volume} {114}},\
  \bibinfo {pages} {237001} (\bibinfo {year} {2015})}\BibitemShut {NoStop}%
\bibitem [{\citenamefont {Ruan}\ \emph
  {et~al.}(2016{\natexlab{a}})\citenamefont {Ruan}, \citenamefont {Jian},
  \citenamefont {Yao}, \citenamefont {Zhang}, \citenamefont {Zhang},\ and\
  \citenamefont {Xing}}]{ruan2016a}%
  \BibitemOpen
  \bibfield  {author} {\bibinfo {author} {\bibfnamefont {J.}~\bibnamefont
  {Ruan}}, \bibinfo {author} {\bibfnamefont {S.-K.}\ \bibnamefont {Jian}},
  \bibinfo {author} {\bibfnamefont {H.}~\bibnamefont {Yao}}, \bibinfo {author}
  {\bibfnamefont {H.}~\bibnamefont {Zhang}}, \bibinfo {author} {\bibfnamefont
  {S.-C.}\ \bibnamefont {Zhang}},\ and\ \bibinfo {author} {\bibfnamefont
  {D.}~\bibnamefont {Xing}},\ }\bibfield  {title} {\bibinfo {title}
  {Symmetry-protected ideal weyl semimetal in hgte-class materials},\ }\href
  {https://www.nature.com/articles/ncomms11136} {\bibfield  {journal} {\bibinfo
   {journal} {Nature communications}\ }\textbf {\bibinfo {volume} {7}},\
  \bibinfo {pages} {1} (\bibinfo {year} {2016}{\natexlab{a}})}\BibitemShut
  {NoStop}%
\bibitem [{\citenamefont {Ruan}\ \emph
  {et~al.}(2016{\natexlab{b}})\citenamefont {Ruan}, \citenamefont {Jian},
  \citenamefont {Zhang}, \citenamefont {Yao}, \citenamefont {Zhang},
  \citenamefont {Zhang},\ and\ \citenamefont {Xing}}]{ruan2016b}%
  \BibitemOpen
  \bibfield  {author} {\bibinfo {author} {\bibfnamefont {J.}~\bibnamefont
  {Ruan}}, \bibinfo {author} {\bibfnamefont {S.-K.}\ \bibnamefont {Jian}},
  \bibinfo {author} {\bibfnamefont {D.}~\bibnamefont {Zhang}}, \bibinfo
  {author} {\bibfnamefont {H.}~\bibnamefont {Yao}}, \bibinfo {author}
  {\bibfnamefont {H.}~\bibnamefont {Zhang}}, \bibinfo {author} {\bibfnamefont
  {S.-C.}\ \bibnamefont {Zhang}},\ and\ \bibinfo {author} {\bibfnamefont
  {D.}~\bibnamefont {Xing}},\ }\bibfield  {title} {\bibinfo {title} {Ideal weyl
  semimetals in the chalcopyrites ${\mathrm{cutlse}}_{2}$,
  ${\mathrm{agtlte}}_{2}$, ${\mathrm{autlte}}_{2}$, and
  ${\mathrm{znpbas}}_{2}$},\ }\href
  {https://doi.org/10.1103/PhysRevLett.116.226801} {\bibfield  {journal}
  {\bibinfo  {journal} {Phys. Rev. Lett.}\ }\textbf {\bibinfo {volume} {116}},\
  \bibinfo {pages} {226801} (\bibinfo {year} {2016}{\natexlab{b}})}\BibitemShut
  {NoStop}%
\bibitem [{\citenamefont {Lee}\ and\ \citenamefont
  {Ramakrishnan}(1985)}]{Lee85}%
  \BibitemOpen
  \bibfield  {author} {\bibinfo {author} {\bibfnamefont {P.~A.}\ \bibnamefont
  {Lee}}\ and\ \bibinfo {author} {\bibfnamefont {T.~V.}\ \bibnamefont
  {Ramakrishnan}},\ }\bibfield  {title} {\bibinfo {title} {Disordered
  electronic systems},\ }\href {https://doi.org/10.1103/RevModPhys.57.287}
  {\bibfield  {journal} {\bibinfo  {journal} {Rev. Mod. Phys.}\ }\textbf
  {\bibinfo {volume} {57}},\ \bibinfo {pages} {287} (\bibinfo {year}
  {1985})}\BibitemShut {NoStop}%
\bibitem [{\citenamefont {Altshuler}\ and\ \citenamefont
  {Aronov}(1985)}]{Altshuler}%
  \BibitemOpen
  \bibfield  {author} {\bibinfo {author} {\bibfnamefont {B.}~\bibnamefont
  {Altshuler}}\ and\ \bibinfo {author} {\bibfnamefont {A.~G.}\ \bibnamefont
  {Aronov}},\ }\bibfield  {title} {\bibinfo {title} {Electron-electron
  interactions in disordered systems ed al efros and m pollak},\ }\href@noop {}
  {\bibfield  {journal} {\bibinfo  {journal} {Amsterdam: North-Holland) p}\
  }\textbf {\bibinfo {volume} {1}},\ \bibinfo {pages} {155} (\bibinfo {year}
  {1985})}\BibitemShut {NoStop}%
\bibitem [{\citenamefont {Fradkin}(1986)}]{Fradkin1986}%
  \BibitemOpen
  \bibfield  {author} {\bibinfo {author} {\bibfnamefont {E.}~\bibnamefont
  {Fradkin}},\ }\bibfield  {title} {\bibinfo {title} {Critical behavior of
  disordered degenerate semiconductors. ii. spectrum and transport properties
  in mean-field theory},\ }\href {https://doi.org/10.1103/PhysRevB.33.3263}
  {\bibfield  {journal} {\bibinfo  {journal} {Phys. Rev. B}\ }\textbf {\bibinfo
  {volume} {33}},\ \bibinfo {pages} {3263} (\bibinfo {year}
  {1986})}\BibitemShut {NoStop}%
\bibitem [{\citenamefont {Belitz}\ and\ \citenamefont
  {Kirkpatrick}(1994)}]{Belitz94}%
  \BibitemOpen
  \bibfield  {author} {\bibinfo {author} {\bibfnamefont {D.}~\bibnamefont
  {Belitz}}\ and\ \bibinfo {author} {\bibfnamefont {T.~R.}\ \bibnamefont
  {Kirkpatrick}},\ }\bibfield  {title} {\bibinfo {title} {The anderson-mott
  transition},\ }\href {https://doi.org/10.1103/RevModPhys.66.261} {\bibfield
  {journal} {\bibinfo  {journal} {Rev. Mod. Phys.}\ }\textbf {\bibinfo {volume}
  {66}},\ \bibinfo {pages} {261} (\bibinfo {year} {1994})}\BibitemShut
  {NoStop}%
\bibitem [{\citenamefont {Abrahams}\ \emph {et~al.}(2001)\citenamefont
  {Abrahams}, \citenamefont {Kravchenko},\ and\ \citenamefont
  {Sarachik}}]{Abrahams01}%
  \BibitemOpen
  \bibfield  {author} {\bibinfo {author} {\bibfnamefont {E.}~\bibnamefont
  {Abrahams}}, \bibinfo {author} {\bibfnamefont {S.~V.}\ \bibnamefont
  {Kravchenko}},\ and\ \bibinfo {author} {\bibfnamefont {M.~P.}\ \bibnamefont
  {Sarachik}},\ }\bibfield  {title} {\bibinfo {title} {Metallic behavior and
  related phenomena in two dimensions},\ }\href
  {https://doi.org/10.1103/RevModPhys.73.251} {\bibfield  {journal} {\bibinfo
  {journal} {Rev. Mod. Phys.}\ }\textbf {\bibinfo {volume} {73}},\ \bibinfo
  {pages} {251} (\bibinfo {year} {2001})}\BibitemShut {NoStop}%
\bibitem [{\citenamefont {Altland}\ \emph {et~al.}(2002)\citenamefont
  {Altland}, \citenamefont {Simons},\ and\ \citenamefont
  {Zirnbauer}}]{Altland02}%
  \BibitemOpen
  \bibfield  {author} {\bibinfo {author} {\bibfnamefont {A.}~\bibnamefont
  {Altland}}, \bibinfo {author} {\bibfnamefont {B.}~\bibnamefont {Simons}},\
  and\ \bibinfo {author} {\bibfnamefont {M.}~\bibnamefont {Zirnbauer}},\
  }\bibfield  {title} {\bibinfo {title} {Theories of low-energy quasi-particle
  states in disordered d-wave superconductors},\ }\href
  {https://doi.org/https://doi.org/10.1016/S0370-1573(01)00065-5} {\bibfield
  {journal} {\bibinfo  {journal} {Physics Reports}\ }\textbf {\bibinfo {volume}
  {359}},\ \bibinfo {pages} {283} (\bibinfo {year} {2002})}\BibitemShut
  {NoStop}%
\bibitem [{\citenamefont {Das~Sarma}\ \emph {et~al.}(2011)\citenamefont
  {Das~Sarma}, \citenamefont {Adam}, \citenamefont {Hwang},\ and\ \citenamefont
  {Rossi}}]{Sarma11}%
  \BibitemOpen
  \bibfield  {author} {\bibinfo {author} {\bibfnamefont {S.}~\bibnamefont
  {Das~Sarma}}, \bibinfo {author} {\bibfnamefont {S.}~\bibnamefont {Adam}},
  \bibinfo {author} {\bibfnamefont {E.~H.}\ \bibnamefont {Hwang}},\ and\
  \bibinfo {author} {\bibfnamefont {E.}~\bibnamefont {Rossi}},\ }\bibfield
  {title} {\bibinfo {title} {Electronic transport in two-dimensional
  graphene},\ }\href {https://doi.org/10.1103/RevModPhys.83.407} {\bibfield
  {journal} {\bibinfo  {journal} {Rev. Mod. Phys.}\ }\textbf {\bibinfo {volume}
  {83}},\ \bibinfo {pages} {407} (\bibinfo {year} {2011})}\BibitemShut
  {NoStop}%
\bibitem [{\citenamefont {Kotov}\ \emph {et~al.}(2012)\citenamefont {Kotov},
  \citenamefont {Uchoa}, \citenamefont {Pereira}, \citenamefont {Guinea},\ and\
  \citenamefont {Castro~Neto}}]{Kotov12}%
  \BibitemOpen
  \bibfield  {author} {\bibinfo {author} {\bibfnamefont {V.~N.}\ \bibnamefont
  {Kotov}}, \bibinfo {author} {\bibfnamefont {B.}~\bibnamefont {Uchoa}},
  \bibinfo {author} {\bibfnamefont {V.~M.}\ \bibnamefont {Pereira}}, \bibinfo
  {author} {\bibfnamefont {F.}~\bibnamefont {Guinea}},\ and\ \bibinfo {author}
  {\bibfnamefont {A.~H.}\ \bibnamefont {Castro~Neto}},\ }\bibfield  {title}
  {\bibinfo {title} {Electron-electron interactions in graphene: Current status
  and perspectives},\ }\href {https://doi.org/10.1103/RevModPhys.84.1067}
  {\bibfield  {journal} {\bibinfo  {journal} {Rev. Mod. Phys.}\ }\textbf
  {\bibinfo {volume} {84}},\ \bibinfo {pages} {1067} (\bibinfo {year}
  {2012})}\BibitemShut {NoStop}%
\bibitem [{\citenamefont {Yerzhakov}\ and\ \citenamefont
  {Maciejko}(2018)}]{yerzhakov2018prb}%
  \BibitemOpen
  \bibfield  {author} {\bibinfo {author} {\bibfnamefont {H.}~\bibnamefont
  {Yerzhakov}}\ and\ \bibinfo {author} {\bibfnamefont {J.}~\bibnamefont
  {Maciejko}},\ }\bibfield  {title} {\bibinfo {title} {Disordered fermionic
  quantum critical points},\ }\href
  {https://doi.org/10.1103/PhysRevB.98.195142} {\bibfield  {journal} {\bibinfo
  {journal} {Phys. Rev. B}\ }\textbf {\bibinfo {volume} {98}},\ \bibinfo
  {pages} {195142} (\bibinfo {year} {2018})}\BibitemShut {NoStop}%
\bibitem [{\citenamefont {Stauber}\ \emph {et~al.}(2005)\citenamefont
  {Stauber}, \citenamefont {Guinea},\ and\ \citenamefont
  {Vozmediano}}]{Stauber2005PRB}%
  \BibitemOpen
  \bibfield  {author} {\bibinfo {author} {\bibfnamefont {T.}~\bibnamefont
  {Stauber}}, \bibinfo {author} {\bibfnamefont {F.}~\bibnamefont {Guinea}},\
  and\ \bibinfo {author} {\bibfnamefont {M.~A.~H.}\ \bibnamefont
  {Vozmediano}},\ }\bibfield  {title} {\bibinfo {title} {Disorder and
  interaction effects in two-dimensional graphene sheets},\ }\href
  {https://doi.org/10.1103/PhysRevB.71.041406} {\bibfield  {journal} {\bibinfo
  {journal} {Phys. Rev. B}\ }\textbf {\bibinfo {volume} {71}},\ \bibinfo
  {pages} {041406} (\bibinfo {year} {2005})}\BibitemShut {NoStop}%
\bibitem [{\citenamefont {Herbut}\ \emph {et~al.}(2008)\citenamefont {Herbut},
  \citenamefont {Juri\ifmmode \check{c}\else \v{c}\fi{}i\ifmmode~\acute{c}\else
  \'{c}\fi{}},\ and\ \citenamefont {Vafek}}]{Herbut08}%
  \BibitemOpen
  \bibfield  {author} {\bibinfo {author} {\bibfnamefont {I.~F.}\ \bibnamefont
  {Herbut}}, \bibinfo {author} {\bibfnamefont {V.}~\bibnamefont {Juri\ifmmode
  \check{c}\else \v{c}\fi{}i\ifmmode~\acute{c}\else \'{c}\fi{}}},\ and\
  \bibinfo {author} {\bibfnamefont {O.}~\bibnamefont {Vafek}},\ }\bibfield
  {title} {\bibinfo {title} {Coulomb interaction, ripples, and the minimal
  conductivity of graphene},\ }\href
  {https://doi.org/10.1103/PhysRevLett.100.046403} {\bibfield  {journal}
  {\bibinfo  {journal} {Phys. Rev. Lett.}\ }\textbf {\bibinfo {volume} {100}},\
  \bibinfo {pages} {046403} (\bibinfo {year} {2008})}\BibitemShut {NoStop}%
\bibitem [{\citenamefont {Vafek}\ and\ \citenamefont {Case}(2008)}]{Vafek08}%
  \BibitemOpen
  \bibfield  {author} {\bibinfo {author} {\bibfnamefont {O.}~\bibnamefont
  {Vafek}}\ and\ \bibinfo {author} {\bibfnamefont {M.~J.}\ \bibnamefont
  {Case}},\ }\bibfield  {title} {\bibinfo {title} {Renormalization group
  approach to two-dimensional coulomb interacting dirac fermions with random
  gauge potential},\ }\href {https://doi.org/10.1103/PhysRevB.77.033410}
  {\bibfield  {journal} {\bibinfo  {journal} {Phys. Rev. B}\ }\textbf {\bibinfo
  {volume} {77}},\ \bibinfo {pages} {033410} (\bibinfo {year}
  {2008})}\BibitemShut {NoStop}%
\bibitem [{\citenamefont {Goswami}\ and\ \citenamefont
  {Chakravarty}(2011)}]{Goswami2011PRL}%
  \BibitemOpen
  \bibfield  {author} {\bibinfo {author} {\bibfnamefont {P.}~\bibnamefont
  {Goswami}}\ and\ \bibinfo {author} {\bibfnamefont {S.}~\bibnamefont
  {Chakravarty}},\ }\bibfield  {title} {\bibinfo {title} {Quantum criticality
  between topological and band insulators in $3+1$ dimensions},\ }\href
  {https://doi.org/10.1103/PhysRevLett.107.196803} {\bibfield  {journal}
  {\bibinfo  {journal} {Phys. Rev. Lett.}\ }\textbf {\bibinfo {volume} {107}},\
  \bibinfo {pages} {196803} (\bibinfo {year} {2011})}\BibitemShut {NoStop}%
\bibitem [{\citenamefont {Wang}\ and\ \citenamefont {Liu}(2014)}]{WangLiu14}%
  \BibitemOpen
  \bibfield  {author} {\bibinfo {author} {\bibfnamefont {J.-R.}\ \bibnamefont
  {Wang}}\ and\ \bibinfo {author} {\bibfnamefont {G.-Z.}\ \bibnamefont {Liu}},\
  }\bibfield  {title} {\bibinfo {title} {Influence of coulomb interaction on
  the anisotropic dirac cone in graphene},\ }\href
  {https://doi.org/10.1103/PhysRevB.89.195404} {\bibfield  {journal} {\bibinfo
  {journal} {Phys. Rev. B}\ }\textbf {\bibinfo {volume} {89}},\ \bibinfo
  {pages} {195404} (\bibinfo {year} {2014})}\BibitemShut {NoStop}%
\bibitem [{\citenamefont {Roy}\ and\ \citenamefont
  {Das~Sarma}(2014)}]{Roy2014PRB}%
  \BibitemOpen
  \bibfield  {author} {\bibinfo {author} {\bibfnamefont {B.}~\bibnamefont
  {Roy}}\ and\ \bibinfo {author} {\bibfnamefont {S.}~\bibnamefont
  {Das~Sarma}},\ }\bibfield  {title} {\bibinfo {title} {Diffusive quantum
  criticality in three-dimensional disordered dirac semimetals},\ }\href
  {https://doi.org/10.1103/PhysRevB.90.241112} {\bibfield  {journal} {\bibinfo
  {journal} {Phys. Rev. B}\ }\textbf {\bibinfo {volume} {90}},\ \bibinfo
  {pages} {241112} (\bibinfo {year} {2014})}\BibitemShut {NoStop}%
\bibitem [{\citenamefont {Zhao}\ \emph {et~al.}(2016)\citenamefont {Zhao},
  \citenamefont {Wang}, \citenamefont {Wang},\ and\ \citenamefont
  {Liu}}]{ZhaoPRB2016}%
  \BibitemOpen
  \bibfield  {author} {\bibinfo {author} {\bibfnamefont {P.-L.}\ \bibnamefont
  {Zhao}}, \bibinfo {author} {\bibfnamefont {J.-R.}\ \bibnamefont {Wang}},
  \bibinfo {author} {\bibfnamefont {A.-M.}\ \bibnamefont {Wang}},\ and\
  \bibinfo {author} {\bibfnamefont {G.-Z.}\ \bibnamefont {Liu}},\ }\bibfield
  {title} {\bibinfo {title} {Interplay of coulomb interaction and disorder in a
  two-dimensional semi-dirac fermion system},\ }\href
  {https://doi.org/10.1103/PhysRevB.94.195114} {\bibfield  {journal} {\bibinfo
  {journal} {Phys. Rev. B}\ }\textbf {\bibinfo {volume} {94}},\ \bibinfo
  {pages} {195114} (\bibinfo {year} {2016})}\BibitemShut {NoStop}%
\bibitem [{\citenamefont {Ostrovsky}\ \emph {et~al.}(2006)\citenamefont
  {Ostrovsky}, \citenamefont {Gornyi},\ and\ \citenamefont
  {Mirlin}}]{Ostrovsky06}%
  \BibitemOpen
  \bibfield  {author} {\bibinfo {author} {\bibfnamefont {P.~M.}\ \bibnamefont
  {Ostrovsky}}, \bibinfo {author} {\bibfnamefont {I.~V.}\ \bibnamefont
  {Gornyi}},\ and\ \bibinfo {author} {\bibfnamefont {A.~D.}\ \bibnamefont
  {Mirlin}},\ }\bibfield  {title} {\bibinfo {title} {Electron transport in
  disordered graphene},\ }\href {https://doi.org/10.1103/PhysRevB.74.235443}
  {\bibfield  {journal} {\bibinfo  {journal} {Phys. Rev. B}\ }\textbf {\bibinfo
  {volume} {74}},\ \bibinfo {pages} {235443} (\bibinfo {year}
  {2006})}\BibitemShut {NoStop}%
\bibitem [{\citenamefont {Foster}\ and\ \citenamefont
  {Aleiner}(2008)}]{Foster08}%
  \BibitemOpen
  \bibfield  {author} {\bibinfo {author} {\bibfnamefont {M.~S.}\ \bibnamefont
  {Foster}}\ and\ \bibinfo {author} {\bibfnamefont {I.~L.}\ \bibnamefont
  {Aleiner}},\ }\bibfield  {title} {\bibinfo {title} {Graphene via large $n$: A
  renormalization group study},\ }\href
  {https://doi.org/10.1103/PhysRevB.77.195413} {\bibfield  {journal} {\bibinfo
  {journal} {Phys. Rev. B}\ }\textbf {\bibinfo {volume} {77}},\ \bibinfo
  {pages} {195413} (\bibinfo {year} {2008})}\BibitemShut {NoStop}%
\bibitem [{\citenamefont {Yerzhakov}\ and\ \citenamefont
  {Maciejko}(2021)}]{yerzhakov2020npb}%
  \BibitemOpen
  \bibfield  {author} {\bibinfo {author} {\bibfnamefont {H.}~\bibnamefont
  {Yerzhakov}}\ and\ \bibinfo {author} {\bibfnamefont {J.}~\bibnamefont
  {Maciejko}},\ }\bibfield  {title} {\bibinfo {title} {Random-mass disorder in
  the critical gross-neveu-yukawa models},\ }\href
  {https://doi.org/https://doi.org/10.1016/j.nuclphysb.2020.115241} {\bibfield
  {journal} {\bibinfo  {journal} {Nuclear Physics B}\ }\textbf {\bibinfo
  {volume} {962}},\ \bibinfo {pages} {115241} (\bibinfo {year}
  {2021})}\BibitemShut {NoStop}%
\bibitem [{\citenamefont {Roy}\ \emph {et~al.}(2016)\citenamefont {Roy},
  \citenamefont {Juri{\v{c}}i{\'c}},\ and\ \citenamefont
  {Das~Sarma}}]{Roy2016SCP}%
  \BibitemOpen
  \bibfield  {author} {\bibinfo {author} {\bibfnamefont {B.}~\bibnamefont
  {Roy}}, \bibinfo {author} {\bibfnamefont {V.}~\bibnamefont
  {Juri{\v{c}}i{\'c}}},\ and\ \bibinfo {author} {\bibfnamefont
  {S.}~\bibnamefont {Das~Sarma}},\ }\bibfield  {title} {\bibinfo {title}
  {Universal optical conductivity of a disordered weyl semimetal},\ }\href
  {https://www.nature.com/articles/srep32446} {\bibfield  {journal} {\bibinfo
  {journal} {Scientific reports}\ }\textbf {\bibinfo {volume} {6}},\ \bibinfo
  {pages} {1} (\bibinfo {year} {2016})}\BibitemShut {NoStop}%
\bibitem [{\citenamefont {Kim}\ \emph {et~al.}(1997)\citenamefont {Kim},
  \citenamefont {Lee},\ and\ \citenamefont {Wen}}]{Kim97}%
  \BibitemOpen
  \bibfield  {author} {\bibinfo {author} {\bibfnamefont {D.~H.}\ \bibnamefont
  {Kim}}, \bibinfo {author} {\bibfnamefont {P.~A.}\ \bibnamefont {Lee}},\ and\
  \bibinfo {author} {\bibfnamefont {X.-G.}\ \bibnamefont {Wen}},\ }\bibfield
  {title} {\bibinfo {title} {Massless dirac fermions, gauge fields, and
  underdoped cuprates},\ }\href {https://doi.org/10.1103/PhysRevLett.79.2109}
  {\bibfield  {journal} {\bibinfo  {journal} {Phys. Rev. Lett.}\ }\textbf
  {\bibinfo {volume} {79}},\ \bibinfo {pages} {2109} (\bibinfo {year}
  {1997})}\BibitemShut {NoStop}%
\bibitem [{\citenamefont {Lee}\ \emph {et~al.}(2006)\citenamefont {Lee},
  \citenamefont {Nagaosa},\ and\ \citenamefont {Wen}}]{Lee06}%
  \BibitemOpen
  \bibfield  {author} {\bibinfo {author} {\bibfnamefont {P.~A.}\ \bibnamefont
  {Lee}}, \bibinfo {author} {\bibfnamefont {N.}~\bibnamefont {Nagaosa}},\ and\
  \bibinfo {author} {\bibfnamefont {X.-G.}\ \bibnamefont {Wen}},\ }\bibfield
  {title} {\bibinfo {title} {Doping a mott insulator: Physics of
  high-temperature superconductivity},\ }\href
  {https://doi.org/10.1103/RevModPhys.78.17} {\bibfield  {journal} {\bibinfo
  {journal} {Rev. Mod. Phys.}\ }\textbf {\bibinfo {volume} {78}},\ \bibinfo
  {pages} {17} (\bibinfo {year} {2006})}\BibitemShut {NoStop}%
\bibitem [{\citenamefont {Wang}(2013)}]{JWang13}%
  \BibitemOpen
  \bibfield  {author} {\bibinfo {author} {\bibfnamefont {J.}~\bibnamefont
  {Wang}},\ }\bibfield  {title} {\bibinfo {title} {Velocity renormalization of
  nodal quasiparticles in $d$-wave superconductors},\ }\href
  {https://doi.org/10.1103/PhysRevB.87.054511} {\bibfield  {journal} {\bibinfo
  {journal} {Phys. Rev. B}\ }\textbf {\bibinfo {volume} {87}},\ \bibinfo
  {pages} {054511} (\bibinfo {year} {2013})}\BibitemShut {NoStop}%
\bibitem [{\citenamefont {Wang}\ \emph {et~al.}(2011)\citenamefont {Wang},
  \citenamefont {Liu},\ and\ \citenamefont {Kleinert}}]{Wang2011PRB}%
  \BibitemOpen
  \bibfield  {author} {\bibinfo {author} {\bibfnamefont {J.}~\bibnamefont
  {Wang}}, \bibinfo {author} {\bibfnamefont {G.-Z.}\ \bibnamefont {Liu}},\ and\
  \bibinfo {author} {\bibfnamefont {H.}~\bibnamefont {Kleinert}},\ }\bibfield
  {title} {\bibinfo {title} {Disorder effects at a nematic quantum critical
  point in $d$-wave cuprate superconductors},\ }\href
  {https://doi.org/10.1103/PhysRevB.83.214503} {\bibfield  {journal} {\bibinfo
  {journal} {Phys. Rev. B}\ }\textbf {\bibinfo {volume} {83}},\ \bibinfo
  {pages} {214503} (\bibinfo {year} {2011})}\BibitemShut {NoStop}%
\bibitem [{\citenamefont {Furneaux}\ \emph {et~al.}(1995)\citenamefont
  {Furneaux}, \citenamefont {Kravchenko}, \citenamefont {Mason}, \citenamefont
  {Bowker},\ and\ \citenamefont {Pudalov}}]{Furneaux1995}%
  \BibitemOpen
  \bibfield  {author} {\bibinfo {author} {\bibfnamefont {J.~E.}\ \bibnamefont
  {Furneaux}}, \bibinfo {author} {\bibfnamefont {S.~V.}\ \bibnamefont
  {Kravchenko}}, \bibinfo {author} {\bibfnamefont {W.~E.}\ \bibnamefont
  {Mason}}, \bibinfo {author} {\bibfnamefont {G.~E.}\ \bibnamefont {Bowker}},\
  and\ \bibinfo {author} {\bibfnamefont {V.~M.}\ \bibnamefont {Pudalov}},\
  }\bibfield  {title} {\bibinfo {title} {Destruction of the quantum hall effect
  with increasing disorder},\ }\href
  {https://doi.org/10.1103/PhysRevB.51.17227} {\bibfield  {journal} {\bibinfo
  {journal} {Phys. Rev. B}\ }\textbf {\bibinfo {volume} {51}},\ \bibinfo
  {pages} {17227} (\bibinfo {year} {1995})}\BibitemShut {NoStop}%
\bibitem [{\citenamefont {Ye}\ and\ \citenamefont {Sachdev}(1998)}]{YePRL98}%
  \BibitemOpen
  \bibfield  {author} {\bibinfo {author} {\bibfnamefont {J.}~\bibnamefont
  {Ye}}\ and\ \bibinfo {author} {\bibfnamefont {S.}~\bibnamefont {Sachdev}},\
  }\bibfield  {title} {\bibinfo {title} {Coulomb interactions at quantum hall
  critical points of systems in a periodic potential},\ }\href
  {https://doi.org/10.1103/PhysRevLett.80.5409} {\bibfield  {journal} {\bibinfo
   {journal} {Phys. Rev. Lett.}\ }\textbf {\bibinfo {volume} {80}},\ \bibinfo
  {pages} {5409} (\bibinfo {year} {1998})}\BibitemShut {NoStop}%
\bibitem [{\citenamefont {Ye}(1999)}]{YePRB99}%
  \BibitemOpen
  \bibfield  {author} {\bibinfo {author} {\bibfnamefont {J.}~\bibnamefont
  {Ye}},\ }\bibfield  {title} {\bibinfo {title} {Effects of weak disorders on
  quantum hall critical points},\ }\href
  {https://doi.org/10.1103/PhysRevB.60.8290} {\bibfield  {journal} {\bibinfo
  {journal} {Phys. Rev. B}\ }\textbf {\bibinfo {volume} {60}},\ \bibinfo
  {pages} {8290} (\bibinfo {year} {1999})}\BibitemShut {NoStop}%
\bibitem [{\citenamefont {Ludwig}\ \emph {et~al.}(1994)\citenamefont {Ludwig},
  \citenamefont {Fisher}, \citenamefont {Shankar},\ and\ \citenamefont
  {Grinstein}}]{Ludwig1994}%
  \BibitemOpen
  \bibfield  {author} {\bibinfo {author} {\bibfnamefont {A.~W.~W.}\
  \bibnamefont {Ludwig}}, \bibinfo {author} {\bibfnamefont {M.~P.~A.}\
  \bibnamefont {Fisher}}, \bibinfo {author} {\bibfnamefont {R.}~\bibnamefont
  {Shankar}},\ and\ \bibinfo {author} {\bibfnamefont {G.}~\bibnamefont
  {Grinstein}},\ }\bibfield  {title} {\bibinfo {title} {Integer quantum hall
  transition: An alternative approach and exact results},\ }\href
  {https://doi.org/10.1103/PhysRevB.50.7526} {\bibfield  {journal} {\bibinfo
  {journal} {Phys. Rev. B}\ }\textbf {\bibinfo {volume} {50}},\ \bibinfo
  {pages} {7526} (\bibinfo {year} {1994})}\BibitemShut {NoStop}%
\bibitem [{\citenamefont {Zhao}\ \emph {et~al.}(2017)\citenamefont {Zhao},
  \citenamefont {Wang},\ and\ \citenamefont {Liu}}]{plzhao16}%
  \BibitemOpen
  \bibfield  {author} {\bibinfo {author} {\bibfnamefont {P.-L.}\ \bibnamefont
  {Zhao}}, \bibinfo {author} {\bibfnamefont {A.-M.}\ \bibnamefont {Wang}},\
  and\ \bibinfo {author} {\bibfnamefont {G.-Z.}\ \bibnamefont {Liu}},\
  }\bibfield  {title} {\bibinfo {title} {Effects of random potentials in
  three-dimensional quantum electrodynamics},\ }\href
  {https://doi.org/10.1103/PhysRevB.95.235144} {\bibfield  {journal} {\bibinfo
  {journal} {Phys. Rev. B}\ }\textbf {\bibinfo {volume} {95}},\ \bibinfo
  {pages} {235144} (\bibinfo {year} {2017})}\BibitemShut {NoStop}%
\bibitem [{\citenamefont {González}\ \emph {et~al.}(1993)\citenamefont
  {González}, \citenamefont {Guinea},\ and\ \citenamefont
  {Vozmediano}}]{Gonzalez93}%
  \BibitemOpen
  \bibfield  {author} {\bibinfo {author} {\bibfnamefont {J.}~\bibnamefont
  {González}}, \bibinfo {author} {\bibfnamefont {F.}~\bibnamefont {Guinea}},\
  and\ \bibinfo {author} {\bibfnamefont {M.}~\bibnamefont {Vozmediano}},\
  }\bibfield  {title} {\bibinfo {title} {The electronic spectrum of fullerenes
  from the dirac equation},\ }\href
  {https://doi.org/https://doi.org/10.1016/0550-3213(93)90009-E} {\bibfield
  {journal} {\bibinfo  {journal} {Nuclear Physics B}\ }\textbf {\bibinfo
  {volume} {406}},\ \bibinfo {pages} {771} (\bibinfo {year}
  {1993})}\BibitemShut {NoStop}%
\bibitem [{\citenamefont {González}\ \emph {et~al.}(1994)\citenamefont
  {González}, \citenamefont {Guinea},\ and\ \citenamefont
  {Vozmediano}}]{Gonzalez94}%
  \BibitemOpen
  \bibfield  {author} {\bibinfo {author} {\bibfnamefont {J.}~\bibnamefont
  {González}}, \bibinfo {author} {\bibfnamefont {F.}~\bibnamefont {Guinea}},\
  and\ \bibinfo {author} {\bibfnamefont {M.}~\bibnamefont {Vozmediano}},\
  }\bibfield  {title} {\bibinfo {title} {Non-fermi liquid behavior of electrons
  in the half-filled honeycomb lattice (a renormalization group approach)},\
  }\href {https://doi.org/https://doi.org/10.1016/0550-3213(94)90410-3}
  {\bibfield  {journal} {\bibinfo  {journal} {Nuclear Physics B}\ }\textbf
  {\bibinfo {volume} {424}},\ \bibinfo {pages} {595} (\bibinfo {year}
  {1994})}\BibitemShut {NoStop}%
\bibitem [{\citenamefont {Nersesyan}\ \emph {et~al.}(1995)\citenamefont
  {Nersesyan}, \citenamefont {Tsvelik},\ and\ \citenamefont
  {Wenger}}]{Nersesyan95}%
  \BibitemOpen
  \bibfield  {author} {\bibinfo {author} {\bibfnamefont {A.}~\bibnamefont
  {Nersesyan}}, \bibinfo {author} {\bibfnamefont {A.}~\bibnamefont {Tsvelik}},\
  and\ \bibinfo {author} {\bibfnamefont {F.}~\bibnamefont {Wenger}},\
  }\bibfield  {title} {\bibinfo {title} {Disorder effects in two-dimensional
  fermi systems with conical spectrum: exact results for the density of
  states},\ }\href
  {https://doi.org/https://doi.org/10.1016/0550-3213(95)00002-A} {\bibfield
  {journal} {\bibinfo  {journal} {Nuclear Physics B}\ }\textbf {\bibinfo
  {volume} {438}},\ \bibinfo {pages} {561} (\bibinfo {year}
  {1995})}\BibitemShut {NoStop}%
\bibitem [{\citenamefont {Castro~Neto}\ \emph {et~al.}(2009)\citenamefont
  {Castro~Neto}, \citenamefont {Guinea}, \citenamefont {Peres}, \citenamefont
  {Novoselov},\ and\ \citenamefont {Geim}}]{CastroNeto}%
  \BibitemOpen
  \bibfield  {author} {\bibinfo {author} {\bibfnamefont {A.~H.}\ \bibnamefont
  {Castro~Neto}}, \bibinfo {author} {\bibfnamefont {F.}~\bibnamefont {Guinea}},
  \bibinfo {author} {\bibfnamefont {N.~M.~R.}\ \bibnamefont {Peres}}, \bibinfo
  {author} {\bibfnamefont {K.~S.}\ \bibnamefont {Novoselov}},\ and\ \bibinfo
  {author} {\bibfnamefont {A.~K.}\ \bibnamefont {Geim}},\ }\bibfield  {title}
  {\bibinfo {title} {The electronic properties of graphene},\ }\href
  {https://doi.org/10.1103/RevModPhys.81.109} {\bibfield  {journal} {\bibinfo
  {journal} {Rev. Mod. Phys.}\ }\textbf {\bibinfo {volume} {81}},\ \bibinfo
  {pages} {109} (\bibinfo {year} {2009})}\BibitemShut {NoStop}%
\bibitem [{\citenamefont {Peres}(2010)}]{Peres2010RMP}%
  \BibitemOpen
  \bibfield  {author} {\bibinfo {author} {\bibfnamefont {N.~M.~R.}\
  \bibnamefont {Peres}},\ }\bibfield  {title} {\bibinfo {title} {Colloquium:
  The transport properties of graphene: An introduction},\ }\href
  {https://doi.org/10.1103/RevModPhys.82.2673} {\bibfield  {journal} {\bibinfo
  {journal} {Rev. Mod. Phys.}\ }\textbf {\bibinfo {volume} {82}},\ \bibinfo
  {pages} {2673} (\bibinfo {year} {2010})}\BibitemShut {NoStop}%
\bibitem [{\citenamefont {Mucciolo}\ and\ \citenamefont
  {Lewenkopf}(2010)}]{Mucciolo2010JPCM}%
  \BibitemOpen
  \bibfield  {author} {\bibinfo {author} {\bibfnamefont {E.~R.}\ \bibnamefont
  {Mucciolo}}\ and\ \bibinfo {author} {\bibfnamefont {C.~H.}\ \bibnamefont
  {Lewenkopf}},\ }\bibfield  {title} {\bibinfo {title} {Disorder and electronic
  transport in graphene},\ }\href
  {https://doi.org/10.1088/0953-8984/22/27/273201} {\bibfield  {journal}
  {\bibinfo  {journal} {Journal of Physics: Condensed Matter}\ }\textbf
  {\bibinfo {volume} {22}},\ \bibinfo {pages} {273201} (\bibinfo {year}
  {2010})}\BibitemShut {NoStop}%
\bibitem [{\citenamefont {Meyer}\ \emph {et~al.}(2007)\citenamefont {Meyer},
  \citenamefont {Geim}, \citenamefont {Katsnelson}, \citenamefont {Novoselov},
  \citenamefont {Booth},\ and\ \citenamefont {Roth}}]{Meyer2007Nature}%
  \BibitemOpen
  \bibfield  {author} {\bibinfo {author} {\bibfnamefont {J.~C.}\ \bibnamefont
  {Meyer}}, \bibinfo {author} {\bibfnamefont {A.~K.}\ \bibnamefont {Geim}},
  \bibinfo {author} {\bibfnamefont {M.~I.}\ \bibnamefont {Katsnelson}},
  \bibinfo {author} {\bibfnamefont {K.~S.}\ \bibnamefont {Novoselov}}, \bibinfo
  {author} {\bibfnamefont {T.~J.}\ \bibnamefont {Booth}},\ and\ \bibinfo
  {author} {\bibfnamefont {S.}~\bibnamefont {Roth}},\ }\bibfield  {title}
  {\bibinfo {title} {The structure of suspended graphene sheets},\ }\href
  {https://www.nature.com/articles/nature05545} {\bibfield  {journal} {\bibinfo
   {journal} {Nature}\ }\textbf {\bibinfo {volume} {446}},\ \bibinfo {pages}
  {60} (\bibinfo {year} {2007})}\BibitemShut {NoStop}%
\bibitem [{\citenamefont {Champel}\ and\ \citenamefont
  {Florens}(2010)}]{Champel2010PRB}%
  \BibitemOpen
  \bibfield  {author} {\bibinfo {author} {\bibfnamefont {T.}~\bibnamefont
  {Champel}}\ and\ \bibinfo {author} {\bibfnamefont {S.}~\bibnamefont
  {Florens}},\ }\bibfield  {title} {\bibinfo {title} {High magnetic field
  theory for the local density of states in graphene with smooth arbitrary
  potential landscapes},\ }\href {https://doi.org/10.1103/PhysRevB.82.045421}
  {\bibfield  {journal} {\bibinfo  {journal} {Phys. Rev. B}\ }\textbf {\bibinfo
  {volume} {82}},\ \bibinfo {pages} {045421} (\bibinfo {year}
  {2010})}\BibitemShut {NoStop}%
\bibitem [{\citenamefont {Viola~Kusminskiy}\ \emph {et~al.}(2011)\citenamefont
  {Viola~Kusminskiy}, \citenamefont {Campbell}, \citenamefont {Castro~Neto},\
  and\ \citenamefont {Guinea}}]{Kusminskiy2011PRB}%
  \BibitemOpen
  \bibfield  {author} {\bibinfo {author} {\bibfnamefont {S.}~\bibnamefont
  {Viola~Kusminskiy}}, \bibinfo {author} {\bibfnamefont {D.~K.}\ \bibnamefont
  {Campbell}}, \bibinfo {author} {\bibfnamefont {A.~H.}\ \bibnamefont
  {Castro~Neto}},\ and\ \bibinfo {author} {\bibfnamefont {F.}~\bibnamefont
  {Guinea}},\ }\bibfield  {title} {\bibinfo {title} {Pinning of a
  two-dimensional membrane on top of a patterned substrate: The case of
  graphene},\ }\href {https://doi.org/10.1103/PhysRevB.83.165405} {\bibfield
  {journal} {\bibinfo  {journal} {Phys. Rev. B}\ }\textbf {\bibinfo {volume}
  {83}},\ \bibinfo {pages} {165405} (\bibinfo {year} {2011})}\BibitemShut
  {NoStop}%
\bibitem [{\citenamefont {Lerner}(2003)}]{Lerner}%
  \BibitemOpen
  \bibfield  {author} {\bibinfo {author} {\bibfnamefont {I.~V.}\ \bibnamefont
  {Lerner}},\ }\href {https://doi.org/10.48550/ARXIV.COND-MAT/0307471}
  {\bibinfo {title} {Nonlinear sigma model for normal and superconducting
  systems: A pedestrian approach}} (\bibinfo {year} {2003})\BibitemShut
  {NoStop}%
\bibitem [{\citenamefont {Roy}\ and\ \citenamefont
  {Das~Sarma}(2016)}]{Roy2016PRB}%
  \BibitemOpen
  \bibfield  {author} {\bibinfo {author} {\bibfnamefont {B.}~\bibnamefont
  {Roy}}\ and\ \bibinfo {author} {\bibfnamefont {S.}~\bibnamefont
  {Das~Sarma}},\ }\bibfield  {title} {\bibinfo {title} {Quantum phases of
  interacting electrons in three-dimensional dirty dirac semimetals},\ }\href
  {https://doi.org/10.1103/PhysRevB.94.115137} {\bibfield  {journal} {\bibinfo
  {journal} {Phys. Rev. B}\ }\textbf {\bibinfo {volume} {94}},\ \bibinfo
  {pages} {115137} (\bibinfo {year} {2016})}\BibitemShut {NoStop}%
\bibitem [{\citenamefont {Shankar}(1994)}]{Shankar1994RMP}%
  \BibitemOpen
  \bibfield  {author} {\bibinfo {author} {\bibfnamefont {R.}~\bibnamefont
  {Shankar}},\ }\bibfield  {title} {\bibinfo {title} {Renormalization-group
  approach to interacting fermions},\ }\href
  {https://doi.org/10.1103/RevModPhys.66.129} {\bibfield  {journal} {\bibinfo
  {journal} {Rev. Mod. Phys.}\ }\textbf {\bibinfo {volume} {66}},\ \bibinfo
  {pages} {129} (\bibinfo {year} {1994})}\BibitemShut {NoStop}%
\bibitem [{\citenamefont {Foster}(2012)}]{FosterPRB2012}%
  \BibitemOpen
  \bibfield  {author} {\bibinfo {author} {\bibfnamefont {M.~S.}\ \bibnamefont
  {Foster}},\ }\bibfield  {title} {\bibinfo {title} {Multifractal nature of the
  surface local density of states in three-dimensional topological insulators
  with magnetic and nonmagnetic disorder},\ }\href
  {https://doi.org/10.1103/PhysRevB.85.085122} {\bibfield  {journal} {\bibinfo
  {journal} {Phys. Rev. B}\ }\textbf {\bibinfo {volume} {85}},\ \bibinfo
  {pages} {085122} (\bibinfo {year} {2012})}\BibitemShut {NoStop}%
\bibitem [{\citenamefont {Nandkishore}\ \emph {et~al.}(2013)\citenamefont
  {Nandkishore}, \citenamefont {Maciejko}, \citenamefont {Huse},\ and\
  \citenamefont {Sondhi}}]{Nandkishore13}%
  \BibitemOpen
  \bibfield  {author} {\bibinfo {author} {\bibfnamefont {R.}~\bibnamefont
  {Nandkishore}}, \bibinfo {author} {\bibfnamefont {J.}~\bibnamefont
  {Maciejko}}, \bibinfo {author} {\bibfnamefont {D.~A.}\ \bibnamefont {Huse}},\
  and\ \bibinfo {author} {\bibfnamefont {S.~L.}\ \bibnamefont {Sondhi}},\
  }\bibfield  {title} {\bibinfo {title} {Superconductivity of disordered dirac
  fermions},\ }\href {https://doi.org/10.1103/PhysRevB.87.174511} {\bibfield
  {journal} {\bibinfo  {journal} {Phys. Rev. B}\ }\textbf {\bibinfo {volume}
  {87}},\ \bibinfo {pages} {174511} (\bibinfo {year} {2013})}\BibitemShut
  {NoStop}%
\bibitem [{\citenamefont {Sbierski}\ \emph {et~al.}(2016)\citenamefont
  {Sbierski}, \citenamefont {Decker},\ and\ \citenamefont
  {Brouwer}}]{Sbierski16}%
  \BibitemOpen
  \bibfield  {author} {\bibinfo {author} {\bibfnamefont {B.}~\bibnamefont
  {Sbierski}}, \bibinfo {author} {\bibfnamefont {K.~S.~C.}\ \bibnamefont
  {Decker}},\ and\ \bibinfo {author} {\bibfnamefont {P.~W.}\ \bibnamefont
  {Brouwer}},\ }\bibfield  {title} {\bibinfo {title} {Weyl node with random
  vector potential},\ }\href {https://doi.org/10.1103/PhysRevB.94.220202}
  {\bibfield  {journal} {\bibinfo  {journal} {Phys. Rev. B}\ }\textbf {\bibinfo
  {volume} {94}},\ \bibinfo {pages} {220202} (\bibinfo {year}
  {2016})}\BibitemShut {NoStop}%
\bibitem [{\citenamefont {Syzranov}\ and\ \citenamefont
  {Radzihovsky}(2018)}]{Syzranov}%
  \BibitemOpen
  \bibfield  {author} {\bibinfo {author} {\bibfnamefont {S.~V.}\ \bibnamefont
  {Syzranov}}\ and\ \bibinfo {author} {\bibfnamefont {L.}~\bibnamefont
  {Radzihovsky}},\ }\bibfield  {title} {\bibinfo {title} {High-dimensional
  disorder-driven phenomena in weyl semimetals, semiconductors, and related
  systems},\ }\href
  {https://www.annualreviews.org/doi/abs/10.1146/annurev-conmatphys-033117-054037?casa_token=YVJ7OVFsDRUAAAAA:sUHw1LxtNuYkWBWgsBwx05b2jSg3qWw23FV9G2unecve7dG5XoEttPwqJvXgdNVFU_v-TMKiEWDoBV8}
  {\bibfield  {journal} {\bibinfo  {journal} {Annual Review of Condensed Matter
  Physics}\ }\textbf {\bibinfo {volume} {9}},\ \bibinfo {pages} {35} (\bibinfo
  {year} {2018})}\BibitemShut {NoStop}%
\end{thebibliography}%

\clearpage
\onecolumngrid

\end{document}